\documentclass[prx, twocolumn, amsmath, amssymb, superscriptaddress, showpacs, nofootinbib]{revtex4-1}

\usepackage{amsmath}    
\usepackage{graphicx}   
\usepackage{verbatim}   
\usepackage{color}      
\usepackage[usenames,dvipsnames,svgnames,table]{xcolor}
\usepackage{subfigure}  
\usepackage[hidelinks]{hyperref}   
\usepackage{multirow}
\usepackage{float}
\usepackage{braket}
\usepackage{bbm}
\usepackage{microtype}       
\usepackage{bibunits}
\usepackage{bbold}
\usepackage{booktabs}   
\usepackage{scrextend}

\graphicspath{{images/}}

\newcommand\numberthis{\addtocounter{equation}{1}\tag{\theequation}}
\newcommand{\puttitle}{Frequency Up-Conversion Schemes for Controlling Superconducting Qubits}

\begin{document}
		
\begin{abstract}
High-fidelity control of superconducting qubits requires the generation of microwave-frequency pulses precisely tailored on nanosecond timescales. These pulses are most commonly synthesized by up-converting and superimposing two narrow-band intermediate-frequency signals referred to as the in-phase~(I) and quadrature~(Q) components. While the calibration of their DC-offsets, relative amplitude and phase allows one to cancel unwanted sideband and carrier leakage, this IQ mixing approach suffers from the presence of additional spurious frequency components. Here, we experimentally study an alternative approach based on double frequency conversion, which overcomes this challenge and circumvents the need for IQ-calibration. We find a spurious-free dynamic range of more than 70$\,$dB and compare the quality of pulse generation against a state-of-the-art IQ mixing scheme by performing repeated single-qubit randomized benchmarking on a superconducting qubit.
\end{abstract}
	
	\title{\puttitle}
	\author{Johannes Herrmann}
	\email{johannes.herrmann@phys.ethz.ch}
	\affiliation{Department of Physics, ETH Zurich, CH-8093 Zurich, Switzerland}
	\affiliation{Zurich Instruments AG, CH-8005 Zurich, Switzerland}
	\author{Christoph Hellings}
	\affiliation{Department of Physics, ETH Zurich, CH-8093 Zurich, Switzerland}
	\author{Stefania Lazar}
	\affiliation{Department of Physics, ETH Zurich, CH-8093 Zurich, Switzerland}
	\author{Fabian Pfäffli}
	\affiliation{Zurich Instruments AG, CH-8005 Zurich, Switzerland}
	\author{Florian Haupt}
	\affiliation{Zurich Instruments AG, CH-8005 Zurich, Switzerland}
	\author{Tobias~Thiele}
	\affiliation{Zurich Instruments AG, CH-8005 Zurich, Switzerland}
	\author{Dante Colao Zanuz}
	\affiliation{Department of Physics, ETH Zurich, CH-8093 Zurich, Switzerland}
	\author{Graham J. Norris}
	\affiliation{Department of Physics, ETH Zurich, CH-8093 Zurich, Switzerland}
	\author{Flavio Heer}
	\affiliation{Zurich Instruments AG, CH-8005 Zurich, Switzerland}
	\author{Christopher Eichler}
	\affiliation{Department of Physics, ETH Zurich, CH-8093 Zurich, Switzerland}
	\author{Andreas Wallraff}
	\affiliation{Department of Physics, ETH Zurich, CH-8093 Zurich, Switzerland}
	\affiliation{Quantum Center, ETH Zurich, CH-8093 Zurich, Switzerland}
	
	\date{\today}	
	
	\maketitle
\section{Introduction} 	
\begin{figure*}[t]
	\centering
	\includegraphics[width = 1\textwidth]{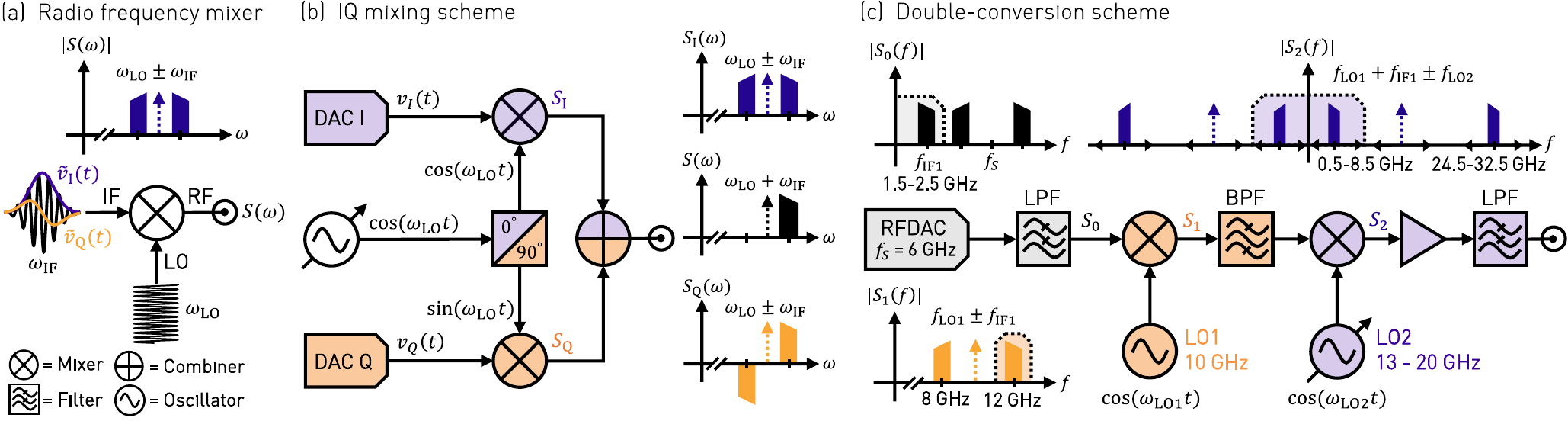}
	\caption{\textbf{Frequency up-conversion schemes.} (a) Diagram and schematic output spectrum $S(\omega)$ of a radio frequency mixer used for up-conversion, where both sidebands at frequencies $\omega_\mathrm{LO}\pm\omega_\mathrm{IF}$ are equally present. The frequency of the LO carrier is indicated with a dashed arrow. (b) Sketch of an IQ mixer, for which the outputs of two parallel up-conversion paths, with an LO signal shared through a 90$^\circ$-hybrid splitter, are superimposed in a microwave combiner. The IF signals for the IQ mixer are generated by two independent digital-to-analog converters (DAC). (c) Sketch of a double frequency conversion scheme with two separate mixing stages fed by two different LO signals. The analog filtering after each stage is indicated by a dashed box in the respective spectrum $S_i(f)$.}
	\label{fig:mix_schemes}
\end{figure*}
Superconducting qubits have emerged as one of the most promising platforms for building error-corrected quantum computers~\cite{Kelly2015, Andersen2020b, Chen2021d, Krinner2021}, which have the potential to solve problems beyond the reach of classical computers~\cite{Preskill2018}. Large-scale quantum computing crucially relies on classical electronics hardware to orchestrate the control and measurement signals. Of particular importance are devices for generating nanosecond-long microwave pulses, which are used for control~\cite{Motzoi2009, Gambetta2011a}, measurement~\cite{Blais2004, Wallraff2005, Walter2017}, and entangling operations~\cite{Chow2011, McKay2017} of superconducting qubits and which also play an important role for operating semiconductor spin qubits~\cite{Malinowski2017a, Barthel2010} and trapped ions~\cite{Harty2016, Ospelkaus2011}.

Microwave pulse generation typically employs a radio frequency mixer to up-convert pulses generated by an arbitrary waveform generator~(AWG) at an intermediate megahertz frequency to the typical gigahertz transition frequency range of the superconducting qubit~\cite{Jolin2020}. Compared to the direct digital synthesis~(DDS) of control pulses~\cite{Raftery2017, Kalfus2020}, frequency up-conversion allows for using AWGs with lower sampling rate, which relaxes resource requirements and thereby eases the scale-up to large channel numbers~\cite{Xu2021d}.

Conventional frequency up-conversion schemes utilize an IQ mixer, which up-converts an in-phase and a quadrature component to destructively interfere unwanted sideband and carrier leakages, which can ultimately limit the achievable single-qubit gate fidelity. However, realistic IQ mixers require extensive calibration of the IQ components to achieve optimal performance, and the up-converted output spectrum exhibits additional spurious frequency components, which cannot be canceled by the interference mechanism.

Here, we control a superconducting qubit with an alternative microwave pulse generation scheme, which makes use of two frequency conversion stages and uses analog filters to remove the unwanted sideband and carrier leakages from the up-converted control pulse. We compare the signal quality of this double frequency conversion scheme to a conventional IQ mixing scheme, and find the output of the double-conversion stage to exhibit smaller spurious frequency components and to be less affected by variations in the ambient temperature.
We also find a slight improvement in the single-qubit gate fidelity when using the double frequency conversion scheme to generate the control pulses.
For our study, we use a high-density IQ converter (HDIQ) and a super-high-frequency signal generator (SHFSG)
to investigate the two schemes.

\section{Frequency Up-Conversion}
High-fidelity single-qubit control is achieved with resonant microwave pulses of the general form
\begin{equation}
V_d(t) = \tilde{v}_{\rm{I}}(t)\cos(\omega t + \phi) + \tilde{v}_{\rm{Q}}(t)\sin(\omega t + \phi),
\label{eq:Vd2}
\end{equation}
where $\omega$ is the transition frequency of the qubit, $\phi$ a global phase, and $\tilde{v}_{\rm{I}}(t)$ and $\tilde{v}_{\rm{Q}}(t)$ are two independent pulse envelope functions. A common choice for $\tilde{v}_{\rm{I}}(t)$ and $\tilde{v}_{\rm{Q}}(t)$ is the Gaussian DRAG pulse parametrization, which allows to avoid leakage into non-computational states~\cite{Motzoi2009, Gambetta2011a, Chen2016}.

To generate the control pulse at frequency~$\omega$ by up-conversion, we multiply an intermediate frequency signal generated by an AWG with a local oscillator continuously running at frequency~$\omega_{\rm{LO}}$ in the gigahertz range. The IF signal consists of the two independent pulse envelope functions $\tilde{v}_{\rm{I}}(t)$ and $\tilde{v}_{\rm{Q}}(t)$ defined in software and modulated digitally at frequency~$\omega_{\rm{IF}}$, see Appendix~A for details. We multiply the IF signal with the LO using a radio frequency mixer resulting in an output spectrum~$S(\omega)$ featuring two sidebands centered around the frequencies~$\omega_{\rm{LO}}\pm\omega_{\rm{IF}}$, see Fig.~\ref{fig:mix_schemes}(a). Throughout the paper, we use the convention of driving the qubit with the upper sideband at frequency $\omega=\omega_{\rm{LO}}+\omega_{\rm{IF}}$ and consider the frequency component $\omega_{\rm{LO}}-\omega_{\rm{IF}}$ the undesired image of the control pulse, which, when present in the spectrum of the qubit drive pulse, induces a gate error. 

\section{IQ mixing scheme}
\label{sec:IQ}
An established method to eliminate the image component is IQ mixing~\cite{Hartley1928, Jolin2020}, as shown in Fig.~\ref{fig:mix_schemes}(b).
Here, two parallel up-conversion paths, for two IF signal waveforms $v_{\rm{I}}(t)$ and $v_{\rm{Q}}(t)$, are equipped with radio frequency~(RF) mixers which share a local oscillator signal through a hybrid splitter which adds a 90$^\circ$-phase shift to the signal driving the Q port mixer. The signals from the two parallel mixing paths are superimposed in a microwave combiner, where ideally one of the two up-converted sidebands is canceled out by destructive interference, see Appendix~A. The spectra in Fig.~\ref{fig:mix_schemes}(b) illustrate this mechanism for the case in which both $v_{\rm{I}}(t)$ and $v_{\rm{Q}}(t)$ have a real-valued Fourier transform. In this particular case, the destructive interference in the microwave combiner is visible from the opposite signs of the lower-sideband signals at frequency $\omega_\mathrm{LO}-\omega_\mathrm{IF}$, see the spectra $S_{\rm{I}}$ and $S_{\rm{Q}}$ after the respective RF mixers in Fig.~\ref{fig:mix_schemes}(b).
To generate the control pulse described by Eq.~\eqref{eq:Vd2} with an IQ mixer, we follow the procedure described in Appendix~A and compute in software two IF signals $v_{\rm{I}}(t)$ and $v_{\rm{Q}}(t)$, which, when up-converted using an ideal IQ mixer, result in an image-free spectrum.

Under realistic conditions, however, the implementation of an IQ mixer is subject to imperfections, such as amplitude and phase imbalances between the two parallel mixing paths~\cite{Jolin2020, Sandia2002}. These imperfections result in an up-converted spectrum that exhibits unwanted sideband and carrier leakage, which we compensate for by calibrating the relative phase, amplitude and DC offsets of the IF signals, as described in the following.

\begin{figure}[t]
	\centering
	\includegraphics[width = 0.48\textwidth]{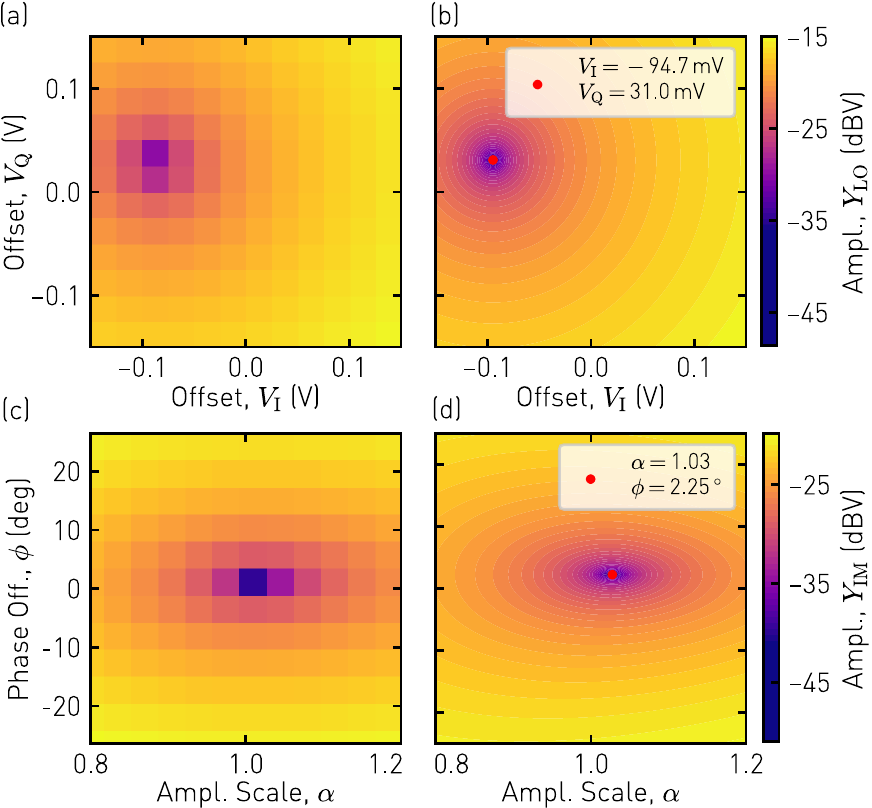}
	\caption{\textbf{IQ mixer calibration.} (a) Measured amplitude $Y_\mathrm{LO}$ of the LO leakage when varying the DC offsets, $V_\mathrm{I}$ and $V_\mathrm{Q}$, and (b) a fit of Eq.~\eqref{eq:LO} to the measured data set. The red data point marks the compensation parameters obtained from the fit. (c) Measured amplitude $Y_\mathrm{IM}$ of the undesired image at frequency  $\omega_\mathrm{LO}-\omega_\mathrm{IF}$ as a function of the amplitude scale $\alpha$ and the phase offset $\phi$ and (d) a fit of Eq.~\eqref{eq:IL} to the measured data set.}		
	\label{fig:iq_calib}
\end{figure}

First, the LO carrier signal leaks through each of the two RF mixers with amplitude $L_\mathrm{I}$, $L_\mathrm{Q}$ and phase $\theta_\mathrm{I}$, $\theta_\mathrm{Q}$, respectively, such that the LO signal at the output of the IQ mixer is~\cite{Sandia2002}
\begin{equation}
y_\mathrm{LO}(t) =  L_\mathrm{I}\cos(\omega_\mathrm{LO}t+\theta_\mathrm{I}) +  L_\mathrm{Q}\sin(\omega_\mathrm{LO}t+\theta_\mathrm{Q}).
\label{eq:LO_V}
\end{equation}
In many basic applications, this effect is not compensated for, but an additional DC bias at the IQ mixer input can help to decrease the amplitude of $y_\mathrm{LO}(t)$. For this purpose, we apply DC voltages $V_\mathrm{I}$ and $V_\mathrm{Q}$ to the IF-ports of the IQ mixer, which create additional signals of the form $V_\mathrm{I}\cos(\omega_\mathrm{LO}t)+ V_\mathrm{Q}\sin(\omega_\mathrm{LO}t)$ at the output. These signals interfere destructively with $y_\mathrm{LO}(t)$ to result in a total amplitude
\begin{align}
Y_\mathrm{LO} = |V_\mathrm{I} + L_\mathrm{I}\mathrm{e}^{\rm i\theta_\mathrm{I}} -\mathrm{i} V_\mathrm{Q} -\mathrm{i} L_\mathrm{Q}\mathrm{e}^{\rm i\theta_\mathrm{Q}}|
\label{eq:LO}
\end{align}
of carrier leakage at the output of the mixer. $Y_\mathrm{LO}$ is minimized for one specific combination of DC compensation voltages.
To find this combination experimentally, we measure the amplitude $Y_\mathrm{LO}$ with an FPGA-based acquisition system, see Appendix~B, while varying the DC offsets $V_{\rm I}$ and $V_{\rm Q}$ with coarse resolution, see Fig.~\ref{fig:iq_calib}(a). We fit Eq.~\eqref{eq:LO} to the measured data to obtain optimal compensation parameters, which are not limited to the resolution between the discrete values of the underlying measurement data and further enhance the calibration accuracy, see Fig.~\ref{fig:iq_calib}(b).

Second, imbalances in the relative amplitude and phase between the two parallel mixing paths modeled by the scaling factor $\tilde\alpha\cdot\rm{e}^{\rm{i}\tilde\phi}$ can result
in an imperfect destructive interference of the image component at frequency $\omega_{\rm{IM}}=\omega_\mathrm{LO}-\omega_\mathrm{IF}$, such that the residual amplitude of the image signal is~\cite{Jolin2020}
\begin{equation}
y_\mathrm{IM}(t) = \frac{1}{2}\cos(\omega_\mathrm{IM}t) - \frac{\tilde\alpha}{2}\cos(\omega_\mathrm{IM}t +\tilde\phi ).
\label{eq:IM}
\end{equation}
We compensate for this imperfection by modifying the IF signals  $v_{\rm{I}}(t)=\cos({\omega_\mathrm{IF}t})$ and $v_{\rm{Q}}=\sin({\omega_\mathrm{IF}t})$ to $v_{\rm{I}}^\prime(t)=\cos({\omega_\mathrm{IF}t})$ and $v_{\rm{Q}}^\prime(t)=\alpha^{-1}\sin({\omega_\mathrm{IF}t}-\phi)$ where $\alpha$ and $\phi$ are the compensation parameters~\cite{Alan2010,Jolin2020} resulting in a total image amplitude
\begin{equation}
Y_\mathrm{IM} = \frac{1}{2} \left|1-\frac{\tilde\alpha}{\alpha}\mathrm{e}^{\rm{i}(\tilde\phi-\phi)}\right|
\label{eq:IL}
\end{equation}
at the output of the mixer, which vanishes for $\tilde\alpha/\alpha=1$ and $\phi-\tilde\phi=2\pi n$ with $n\in \mathbb{N}$. To find the optimal compensation parameters, we measure the amplitude~$Y_{\rm{IM}}$ for varying $\alpha$ and $\phi$, see Fig.~\ref{fig:iq_calib}(c), and fit Eq.~\eqref{eq:IL} to the measured data set similarly as in the calibration procedure of the carrier leakage, see Fig.~\ref{fig:iq_calib}(d).

\section{Double frequency conversion}
\label{sec:double}
An alternative approach to cancel the image component uses analog filtering instead of destructive interference. For conventional RF mixing schemes, however, this would require bandpass filtering with a narrow bandwidth on the order of the IF frequency, which is challenging to realize. In addition, the center frequency of this analog filter would require tunability for addressing qubits in multiple frequency bands. An implementation that overcomes these technical challenges uses a double frequency conversion scheme~\cite{Alan2019}, which makes use of two separate mixing stages fed by two different local oscillators, see Fig.~\ref{fig:mix_schemes}(c). By keeping the frequency of the first LO fixed and having the second LO tunable, a bandpass filter with a fixed center frequency can be used to effectively eliminate sideband and carrier leakage, as described in detail below.

To generate the qubit control signal, we digitally define the control pulse at an intermediate frequency of $f_{\rm{IF1}}=1.5$ to $2.5\,$GHz, see Fig.~\ref{fig:mix_schemes}(c). For that purpose, we employ an RF digital-to-analog converter~(RFDAC) with a sampling rate of $f_{\rm{S}}=6\,$GSa/s resulting in a first Nyquist frequency of $f_{\rm{S}}/2=3\,$GHz. We use a subsequent 2.5$\,$GHz low-pass filter to remove the alias spectra which arise from defining the signal with the RFDAC, and we up-convert the filtered signal with a fixed-frequency LO, which is continuously running at $f_\mathrm{LO1}=10\,$GHz, see spectrum $S_1(f)$ in Fig.~\ref{fig:mix_schemes}(c). We remove the emerging image component at the frequency $f_\mathrm{LO1}-f_{\rm{IF1}}$ using an analog bandpass filter with a pass-band frequency of approximately $12\,$GHz$\,\pm\,500\,$MHz. Then, we down-convert the filtered signal using a second LO with an adjustable frequency range of $f_\mathrm{LO2}=13$ to $20\,$GHz. The image component emerging in the down-converted spectrum ranges from $24.5$ to $32.5\,$GHz and is thus far detuned from the targeted frequency band of $0.5$ to $8.5\,$GHz, see spectrum $S_2(f)$ in Fig.~\ref{fig:mix_schemes}(c). Due to the large detuning, we can use a subsequent low-pass filter to suppress unwanted image components in the qubit control pulse spectrum.

We note that the up-converted spectrum of the qubit control signal is mirrored compared to the spectrum of the initial IF signal since the final down-conversion uses an LO which is higher in frequency than the signal in the previous stage, see also the sketch of the spectrum $S_2(f)$ in Fig.~\ref{fig:mix_schemes}(c). We correct for this effect in software by inverting the sign of the IF frequency $f_\mathrm{IF1}$ prior to the digital up-conversion before the RFDAC.

Compared to the IQ mixing approach, which relies on the destructive interference of two up-converted spectra, the elimination of the image component by analog filtering does not require calibration of the control signals to compensate for RF mixer imperfections. In addition, the carrier leakage of the local oscillators through the RF mixers at $f_\mathrm{LO1}=10\,$GHz and $f_\mathrm{LO2}\geq 12\,$GHz can also effectively be eliminated by means of filtering in the double-conversion scheme.

\section{Verification of signal quality}
\label{sec:spurs}
\subsection{Spurious-free dynamic range}
\begin{figure}[t]
	\centering
	\includegraphics[width = 0.47\textwidth]{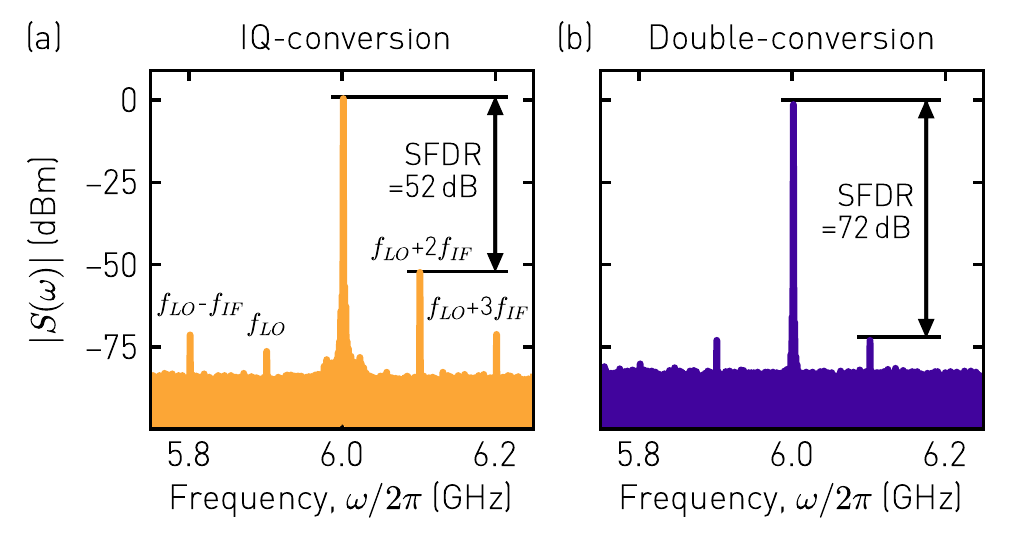}
	\caption{\textbf{Up-converted spectra \& SFDR.} (a) Typical output spectrum $S(\omega)$ of a calibrated IQ mixer and (b)~output spectrum of the double frequency conversion scheme. We indicate the spurious-free dynamic range~(SFDR) in both panels with a double arrow.}
	\label{fig:spec}
\end{figure}

We verify the signal quality of both conversion schemes at room temperature by measuring the corresponding output spectra using an HP8563A spectrum analyzer while generating a continuous wave signal at 6$\,$GHz with a power of 0$\,$dBm. These are typical parameters for achieving fast single-qubit control, see Appendix~B for details.
We then evaluate the spurious-free dynamic range~(SFDR), which is defined as the power ratio between the fundamental signal and the strongest spurious frequency component, see Fig.~\ref{fig:spec}.

We perform IQ mixer calibration, as described in Sec.~\ref{sec:IQ}, and measure the resulting output spectrum~$S(\omega)$ for the input frequencies $f_\mathrm{IF}=100\,$MHz and $f_\mathrm{LO}=5.9\,$GHz.
Here, the level of the LO leakage and the unwanted sideband  are about 70$\,$dB smaller than the desired up-converted signal at $f_\mathrm{LO}+f_\mathrm{IF}=6\,$GHz, see Fig.~\ref{fig:spec}(a). However, the SFDR is limited to 52$\,$dB by a spurious tone at frequency $f_{\rm{LO}}+2f_{\rm{IF}}$. This tone arises from an imperfection of the RF mixer, which under realistic conditions only implements an approximate multiplication of the input signals and additionally creates higher-order polynomial products~\cite{Alan2010, MarkiMixer}. These higher-order products give rise to an output spectrum which contains additional harmonic components at frequencies~\cite{Alan2010}
\begin{equation}
n\cdot f_{\rm{LO}} + m\cdot f_{\rm{IF}} \quad \text{with} \quad n,m \in\mathbb{Z}.
\label{eq:mix}
\end{equation}
Higher order polynomial products with $|n|,|m|\neq 1$ are not suppressed by the interference mechanism in the IQ~mixer and thus degrade the achievable SFDR. 

\begin{figure}[t]
	\centering
	\includegraphics[width = 0.49\textwidth]{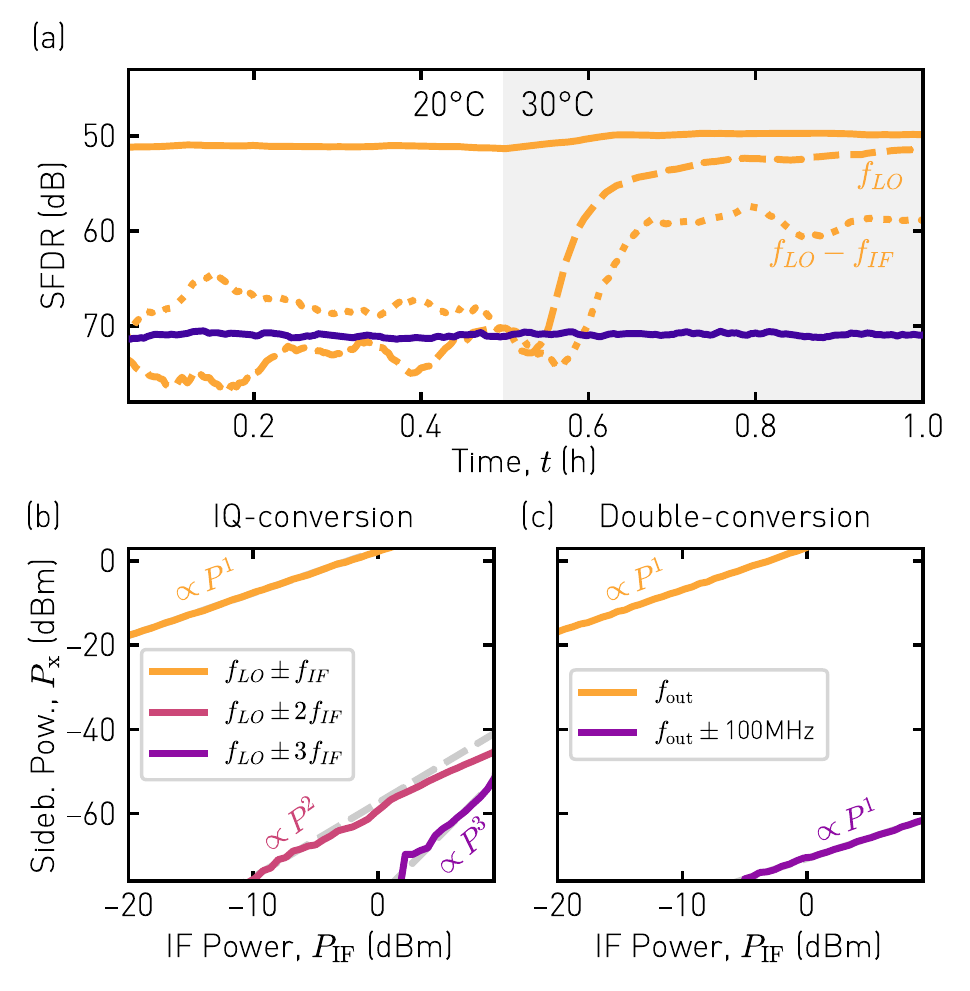}
	\caption{\textbf{Spurious frequency components.} (a)~Response of the SFDR to a step change in ambient temperature for the IQ mixing~(orange) and the double-conversion scheme (purple). The dashed (dotted) orange line displays the suppression of LO leakage (image) in the IQ mixing scheme. (b) IF input power dependence of all spurious IQ mixing frequency components from Fig.~\ref{fig:spec}(a). The dashed gray lines depict a fit with a linear function for each of the power dependencies with a slope corresponding to the order~$m$ of the IF harmonic. (c)~IF input power dependence of all spurious frequency components of the double-conversion scheme from Fig.~\ref{fig:spec}(b).}
	\label{fig:mix_lim}
\end{figure}

For the double-conversion scheme, we generate a continuous wave signal at $f_{\rm{IF1}}=2\,$GHz with the RFDAC and use the frequency $f_\mathrm{LO2}=18\,$GHz to synthesize a 6$\,$GHz output signal. Besides the desired signal at 6$\,$GHz, we observe two spurious frequency components at frequencies 6$\pm$0.1$\,$GHz, which result from crosstalk of the 100$\,$MHz reference used in the instrument. In contrast to the IQ mixing scheme, the higher harmonic frequency components from Eq.~\eqref{eq:mix} can effectively be filtered in the double-conversion scheme, and we attain an SFDR of 72$\,$dB.

To study the stability of both frequency conversion schemes, we measure the SFDR when varying the ambient temperature. For this purpose, we install both systems in a controlled temperature chamber and record the SFDR over the course of 1$\,$h when abruptly changing the ambient temperature from 20$\,^\circ$C to 30$\,^\circ$C after 30$\,$minutes, see Fig.~\ref{fig:mix_lim}(a).
For the IQ mixer, we record, in addition to the SFDR~(orange line), the level of unwanted sideband and carrier leakage~(dashed and dotted orange lines). We find that shortly after the ambient temperature is changed, the suppression of both sideband and carrier leakage degrades significantly, while the SFDR imposed by the second IF harmonic at frequency $f_\mathrm{LO}+2\cdot f_\mathrm{IF}$ is only weakly affected. This behavior can be traced back to the temperature dependence of the IQ mixer imperfections, which result in a drift of the optimal IQ calibration parameters, see Eq.~\eqref{eq:LO} and~\eqref{eq:IL}. The SFDR of the double-conversion scheme~(purple), however, is unaffected by the temperature change since it does not rely on a the destructive interference of two signals.

We also investigate the peak intensity of the spurs shown in Fig.~\ref{fig:spec} in dependence on the IF input power $P$. For the IQ mixing scheme, we find a polynomial power dependence~$\propto P^m$ where $m$ is the polynomial order from Eq.~\eqref{eq:mix}, see Fig.~\ref{fig:mix_lim}(b). The power dependence of the 100$\,$MHz reference spurs in the double-conversion scheme is linear with~$\propto P$ such that the SFDR is unaffected.

Due to the polynomial power dependence $\propto P^m$, a general strategy for suppressing higher harmonic frequency components is the reduction of IF input power. To still achieve the desired output power, an RF-amplifier can therefore be added after the IQ mixer. In addition, harmonics of the LO with order $|n|\neq 1$ can be suppressed with a bandpass filter after the RF amplifier. Both measures are implemented for the IQ-conversion module used in this study with a $4$ to $8\,$GHz bandpass filter and an RF amplifier with 21$\,$dB gain.

\begin{figure}[t]
	\centering
	\includegraphics[width = 0.4375\textwidth]{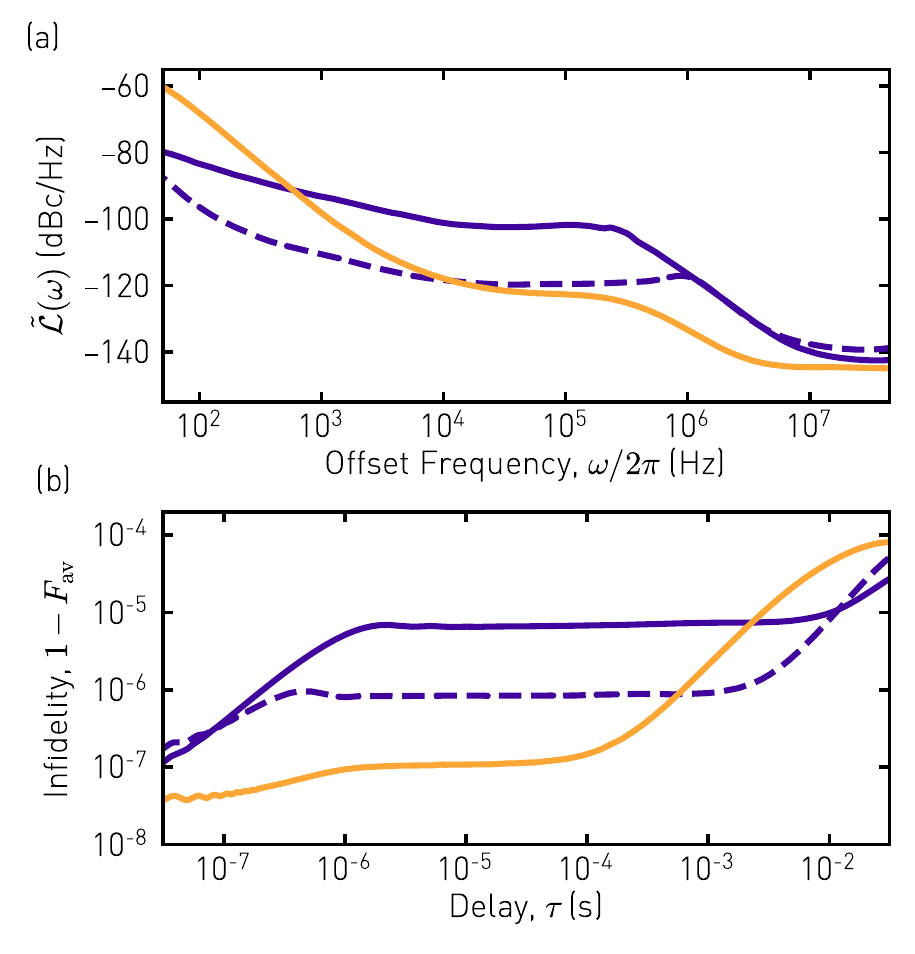}
	\caption{\textbf{Single-sideband phase noise.} (a) Measured single-sideband phase noise $\mathcal{\tilde L}(\omega)$ at 5$\,$GHz for the IQ mixing scheme~(orange), and the double-conversion scheme with~(dashed purple) and without (solid purple) using an LO with improved phase noise. (b)~Process infidelity $1-F_\mathrm{av}$ in a Ramsey experiment with pulse delay $\tau$ calculated based on the noise spectra in (a).}
	\label{fig:pn}
\end{figure}

\subsection{Single-sideband phase noise}
An additional metric of signal quality is the phase noise of the qubit control signal, which can affect the average single-qubit gate fidelity. We measure the single-sideband phase noise $\tilde{\mathcal{L}}(\omega)$ for both frequency conversion schemes at 5$\,$GHz with a signal analyzer~(AnaPico APPH). The orange line in Fig.~\ref{fig:pn}(a) shows the phase noise at the output of the IQ mixer, which is dominated by the noise of the external LO used in the experiment~(R\&S SGS100A).
The phase noise of the signal synthesized with the double-conversion scheme is dictated by the noise of the second LO, which for offset frequencies greater than 1$\,$kHz is larger than the noise from the IQ-conversion module due to the high LO frequency. To estimate the impact of phase noise on gate fidelities, we follow the procedure described in Ref.~\cite{Ball2016b} and compute the infidelity~$1-F_{\mathrm{av}}$  in a Ramsey experiment as a function of the Ramsey delay time~$\tau$. For both frequency conversion schemes, the calculated infidelity is more than one order of magnitude smaller than the limit imposed by qubit coherence, which we determine in the following section, see Fig.~\ref{fig:pn}(b). To assure that phase noise is not a limiting factor once qubit coherence increases, we have improved the circuit design of the LO used in the double frequency conversion scheme, see dashed purple line in Fig.~\ref{fig:pn}, and find a fidelity improvement of about one order of magnitude in the relevant delay regime between micro- and milliseconds.

\section{Verification on quantum hardware}
\label{sec:exp}
To verify the performance of both frequency conversion schemes in controlling quantum hardware, we perform repeated randomized benchmarking~(RB) of single-qubit gates~\cite{Emerson2005, Magesan2011}.

For this purpose, we combine the output of the IQ mixing module~(Zurich Instruments, HDIQ) and of the double frequency conversion module~(Zurich Instruments, SHFSG) with a microwave combiner and connect them to the drive line of a superconducting qubit, which is installed in a cryogenic measurement setup, see Appendix~B for details. For each conversion scheme, we perform standard time-domain calibration of the 50-ns-duration qubit drive pulse using a DRAG pulse parametrization~\cite{Motzoi2009, Gambetta2011a} with a truncated Gaussian envelope. We quantify the amount of leakage into the non-computational second excited state $\ket{f}$ by performing dispersive readout which discriminates between all three states $\ket{g}$, $\ket{e}$, and $\ket{f}$, see Appendix~B.

\begin{figure}[t]
	\centering
	\includegraphics[width = 0.49\textwidth]{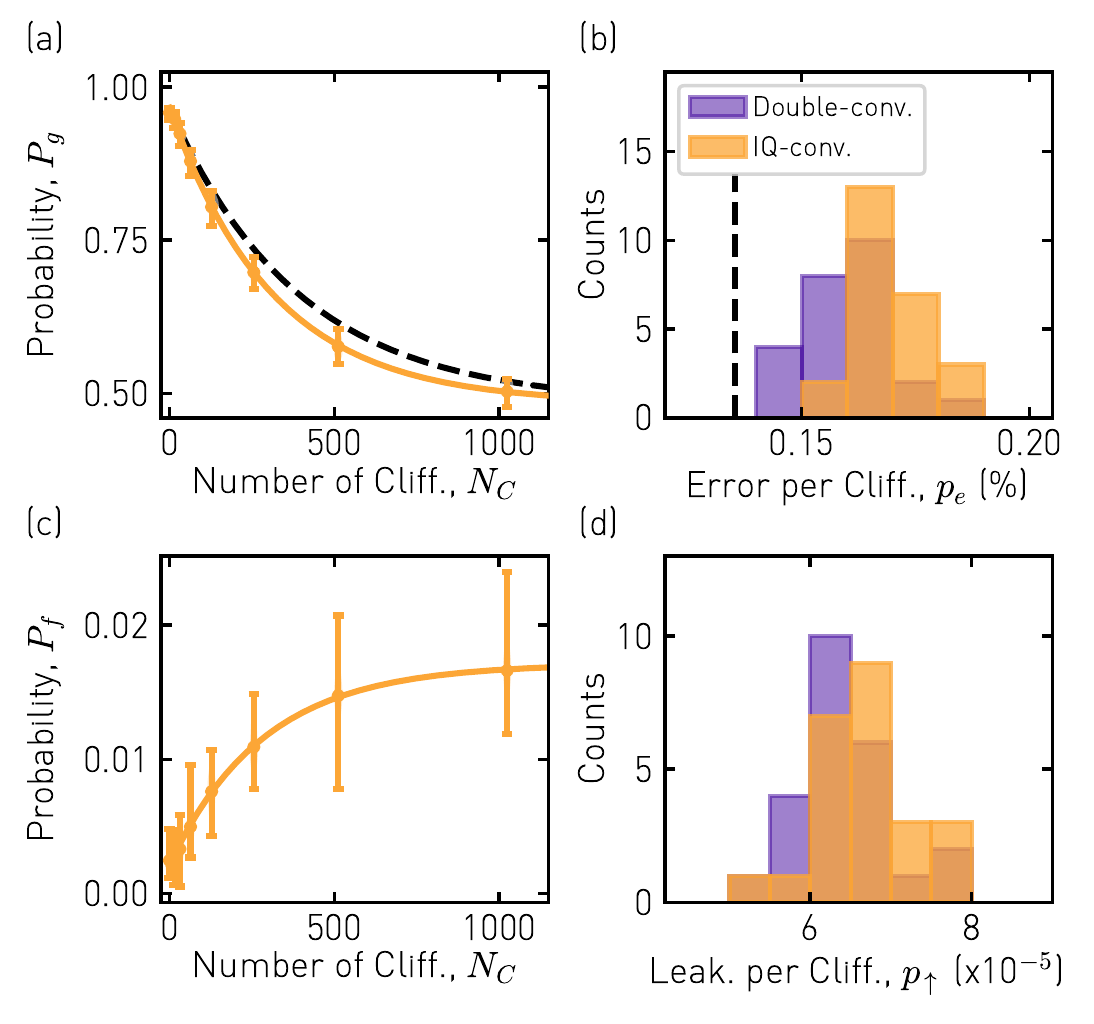}
	\caption{\textbf{Randomized benchmarking.} (a) Measured probability $P_g$ of returning to the ground state \emph{vs.} the number of Clifford gates $N_C$ in a single-qubit randomized benchmarking experiment and estimated coherence limit (dashed black line). (b) Histogram of the measured error per Clifford for 25 repeated single-qubit randomized benchmarking experiments in comparison to the coherence limit (dashed line). (c) Measured leakage probability $P_f$ \emph{vs.} $N_C$ and (d) histogram of the leakage probability per cycle $p_\uparrow$ for the same set of experiments as shown in (b).}
	\label{fig:srb}
\end{figure}

To account for the effect of slow drifts in the system parameters such as the qubit coherence time, we measure the single-qubit gate error repeatedly and alternate~25 times between using the IQ mixer and the double frequency conversion module for applying the RB pulse sequence.
We measure the probability $P_g$ of returning to the ground state after the RB sequence and observe $P_g$ to closely follow the limit imposed by the measured qubit decoherence, which we have calculated according to Ref.~\cite{Asaad2016}, see dashed line in Fig.~\ref{fig:srb}(a) for an example result.

The measured errors per Clifford range from 0.14$\,\%$ to 0.19$\,\%$ in both cases and are close to the coherence limit of 0.13$\,\%$~(dashed line), see histogram of errors obtained from RB in Fig.~\ref{fig:srb}(b). For the IQ mixing scheme, we find an average error per Clifford of $\overline{p}_{e}=0.169\pm 0.007\,\%$, and $0.161\pm 0.008\,\%$ for the double-conversion scheme. In addition to the probability $P_g$ in the RB experiment, we also measure the accumulated leakage~$P_f$ into the non-computational state~$\ket{f}$. By fitting $P_f$ with a rate equation model, as detailed in Ref.~\cite{Chen2016}, we can extract the leakage up-rate $p_\uparrow$ per Clifford, see Fig.~\ref{fig:srb}(c). From the distribution of leakage rates shown in Fig.~\ref{fig:srb}(d), we find an average rate of $\overline{p}_\uparrow=6.8\pm 0.7\times10^{-5}$ for the IQ-conversion scheme and $6.3\pm 0.7\times10^{-5}$ for the double-conversion scheme. For both the measured error and leakage per Clifford, the double-conversion scheme results in values about one standard deviation smaller, which suggests that the double-conversion scheme performs at least as well as the conventional IQ mixing scheme.

\section{Conclusion and Discussion}
By employing a double frequency conversion scheme to synthesize  signals for controlling qubits which are used in superconducting quantum processors, we have demonstrated a promising alternative to the more common IQ mixing approach.
The major advantage of this scheme is an output spectrum with a very low level of spurious frequency components, which does not require extensive calibration and is robust to variations in the laboratory temperature. The benchmarking experiment on quantum hardware has shown a slight improvement in the single-qubit gate fidelity when control pulses are generated with the double-conversion scheme compared to the conventional IQ mixing scheme, which motivates further study of the underlying gate error mechanism.
In addition, the large SFDR of the double-conversion scheme makes it a promising candidate for activating high-fidelity two-qubit gate interactions with RF pulses, such as those used in parametric or cross-resonance gates~\cite{McKay2017, Chow2011}.
The double frequency conversion scheme can also find applications in the readout of superconducting qubits, for which both an up- and down-conversion stage can be realized analogously to the scheme presented in this work.

\section*{Acknowledgments}
We thank Wojciech Ruhnke for contributions to the room temperature measurements, and Michele Collodo and Bruno Küng for input on the manuscript.

The authors acknowledge financial support by ETH Zurich, 
by the EU Flagship on Quantum Technology H2020-FETFLAG2018-03 project 820363 OpenSuperQ, 
by the EU program H2020-FETOPEN project 828826 Quromorphic, 
by the Office of the Director of National Intelligence (ODNI), Intelligence Advanced Research Projects Activity (IARPA), via the U.S. Army Research Office grant W911NF-16-1-0071, 
by the National Center of Competence in Research Quantum Science and Technology (NCCR QSIT), a research instrument of the Swiss National Science Foundation (SNSF) and by the SNFS R’equip grant 206021-170731.
The views and conclusions contained herein are those of the authors and should not be interpreted as necessarily representing the official policies or endorsements, either expressed or implied, of the ODNI, IARPA, or the U.S. Government.

\setcounter{section}{0}
\setcounter{figure}{0}
\setcounter{equation}{0}
\setcounter{table}{0}
\renewcommand*{\thesection}{APPENDIX \Alph{section}}
\renewcommand*{\thefigure}{A\arabic{figure}}
\renewcommand*{\theequation}{A\arabic{equation}}
\renewcommand*{\thetable}{A\arabic{table}}

\section{IQ pulse generation}
\label{Theory}
A generic time-dependent voltage $V_d(t)$, used e.g. for controlling the qubit state, can be parametrized as in Eq.~\eqref{eq:Vd2} as
\begin{equation}
	V_d(t) = \tilde{v}_{\rm{I}}(t)\cos(\omega t + \phi) + \tilde{v}_{\rm{Q}}(t)\sin(\omega t + \phi),
	\label{eq:Vd}
\end{equation}
where $\omega$ is the carrier frequency, $\phi$ is the initial phase of the control voltage, and  $\tilde{v}_{\rm{I}}(t)$ and $\tilde{v}_{\rm{Q}}(t)$ are two independent pulse envelope functions. 

To generate the control voltage $V_d(t)$ from~\eqref{eq:Vd} with the aid of an IQ mixer, we apply to the IF ports the input voltages $v_{\rm{I}}(t)$ and $v_{\rm{Q}}(t)$, which we compute in software according to
\begin{align}
	\begin{pmatrix}
		v_{\rm{I}}   \\
		v_{\rm{Q}}  \\
	\end{pmatrix}
	=
	\begin{pmatrix}
		\cos(\omega_{\rm{IF}}t+\phi) & \sin(\omega_{\rm{IF}}t+\phi)  \\
		-\sin(\omega_{\rm{IF}}t+\phi)   & \cos(\omega_{\rm{IF}}t+\phi)  \\
	\end{pmatrix}
	\begin{pmatrix}
		\tilde{v}_{\rm{I}}   \\
		\tilde{v}_{\rm{Q}}  \\
	\end{pmatrix}
	\label{eq:IQ_MOD}
\end{align}
Using the IQ mixer, the IF signals modulate the continuous wave LO signal such that the signal at the output ideally is
\begin{align}
	V_d(t)
	=
	\begin{pmatrix}
		\cos(\omega_\mathrm{LO}t),\,
		\sin(\omega_\mathrm{LO}t)    \\
	\end{pmatrix}
	\begin{pmatrix}
		v_{\rm{I}}   \\
		v_{\rm{Q}}  \\
	\end{pmatrix}.\label{eq:IQ_MIX}
\end{align}
With the trigonometric identities, $2\cos x\cos y =\cos(x+y) + \cos(x-y)$ and $2\sin x\sin y=\cos(x-y) - \cos(x+y)$, we can simplify~Eq.~\eqref{eq:IQ_MIX} and express the signal at the output of the IQ mixer as
\begin{align*}
	\label{eq:IQ_MIX_OUT}
	V_d(t) &= \tilde{v}_{\rm{I}}(t)\cos\big((\omega_\mathrm{LO}+\omega_\mathrm{IF}) t + \phi\big) \\
	&\quad + \tilde{v}_{\rm{Q}}(t)\sin\big((\omega_\mathrm{LO}+\omega_\mathrm{IF}) t + \phi\big). \numberthis
\end{align*}
When comparing Eq.~\eqref{eq:Vd} to Eq.~\eqref{eq:IQ_MIX_OUT}, it becomes apparent that the output signal is generated at a frequency of $\omega=\omega_{\rm{LO}}+\omega_{\rm{IF}}$. When inverting the sign of the IF frequency $\omega_{\rm{IF}}$ in Eq.~\eqref{eq:IQ_MOD}, the up-converted signal is created at a frequency of~$\omega=\omega_{\rm{LO}}-\omega_{\rm{IF}}$ instead.

\begin{table}[b]
	\begin{tabular}{l c}
		\toprule
		Parameters & Value  \\ \midrule
		Qubit idle frequency, $\omega_\mathrm{q}/2\pi$$\,$(GHz)  &6.143\\
		Qubit anharmonicity, $\alpha_\mathrm{q}/2\pi$$\,$(MHz) &-178  \\
		Lifetime, $T_1$$\,$($\mu$s) & 23.7\\
		Ramsey decay time, $T_2^*$$\,$($\mu$s) & 11.7 \\
		Echo decay time, $T_2^\mathrm{e}$$\,$($\mu$s) & 16.6\\
		Readout resonator frequency, $\omega_\mathrm{r}/2\pi$$\,$(GHz) & 7.141 \\
		Readout linewidth, $\kappa_\mathrm{eff}/2\pi$$\,$(MHz) & 10  \\
		Dispersive Shift, $\chi/2\pi$$\,$(MHz) &  -2.4 \\
		Thermal population, $P_\mathrm{th}$\,$(\%)$ & 3.6\\
		3-level readout assignment prob.$\,$(\%) & 97.2 \\
		\bottomrule
	\end{tabular}
	\caption{Measured parameters of the qubit and of the readout circuit.}
	\label{tab:qb_paras}
\end{table}
\section{Experimental Setup}
\label{setup}	
\subsection{Qubit fabrication and properties} We have fabricated the superconducting transmon qubit in a process similar to the one described in Ref.~\cite{Krinner2021}. Here, we sputter a Niobium thin film onto a high-resistivity intrinsic Silicon substrate and pattern the Niobium layer using photolithography and reactive-ion etching. After patterning the Niobium base layer, we fabricate airbridges to establish well-connected ground planes and
to enable crossings of signal lines. We fabricate Josephson junctions by shadow evaporation of aluminum through a resist mask defined by electron-beam lithography.

We have extracted the qubit and readout circuit parameters, summarized in Table~\ref{tab:qb_paras}, using standard spectroscopy and time-domain measurements.	

\subsection{Wiring and instrumentation} 
\begin{figure}[t]
	\centering
	\includegraphics[width = 0.485\textwidth]{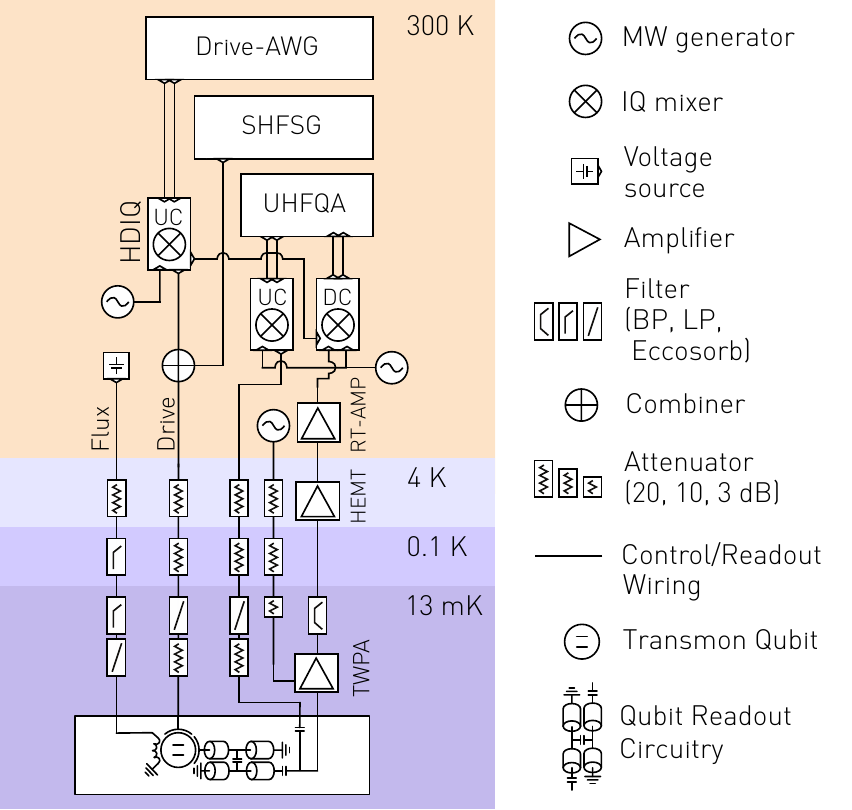}
	\caption{\textbf{Experimental Setup.} Schematic of the control electronics and wiring setup. For details see Appendix~B.}
	\label{fig:setup}
\end{figure}

We have installed the superconducting qubit in a cryogenic measurement setup with a base plate temperature of~13$\,$mK. We have connected the qubit to the control and measurement electronic setup located at room temperature as indicated in Fig.~\ref{fig:setup}. We control the qubit frequency with a magnetic flux, which is generated by a nearby current flowing through a dedicated flux control line. We achieve single-qubit XY-control through a  dedicated drive line, which is capacitively coupled to the qubit and carries the microwave pulses, which are created by the room temperature electronic setup as explained in the main text. As a local oscillator source for the IQ-conversion scheme, we use the Rohde \& Schwarz SGS100A signal generator. To enable the calibration of the IQ mixer, we install an additional switchable RF line which can bypass the cryogenic measurement setup and feed the up-converted spectrum directly into the down-conversion module of the readout line, see Fig.~\ref{fig:setup}.

We perform qubit readout using an FPGA-based measurement system with a sample rate of 1.8$\,$GSa/s~(Zurich Instruments UHFQA). The readout pulse is up-converted to the readout resonator frequency and routed through a highly attenuated~(60$\,$dB) RF line to the qubit chip. After propagating through the readout circuitry of the qubit chip, the readout signal is amplified by a near-quantum-limited traveling wave parametric amplifier~(TWPA)~\cite{Macklin2015}, which is mounted at the base plate of the cryostat. After the TWPA, the readout signal is further amplified by a high-electron mobility transistor~(HEMT) at 4$\,$K, and low-noise amplifiers at room temperature~(RT-AMP). The amplified readout signal is down-converted to an intermediate frequency before being digitized and integrated by the weighted integration units of the UHFQA. 
\subsection{Single-qubit gates} To achieve single-qubit control, we generate microwave pulses with the frequency conversion schemes presented in the main text. We choose the bias of the flux control line to maximize the qubit transition frequency, at which the qubit is to first order insensitive to flux noise, frequently called the upper sweet spot~\cite{Vion2002}. The microwave pulses for single-qubit control follow a Gaussian DRAG pulse parametrization~\cite{Motzoi2009, Gambetta2011a}. with a pulse width of $\sigma=10\,$ns and are truncated at $\pm2.5\sigma=50\,$ns. 
\begin{figure}[t]
	\centering
	\includegraphics[width = 0.42\textwidth]{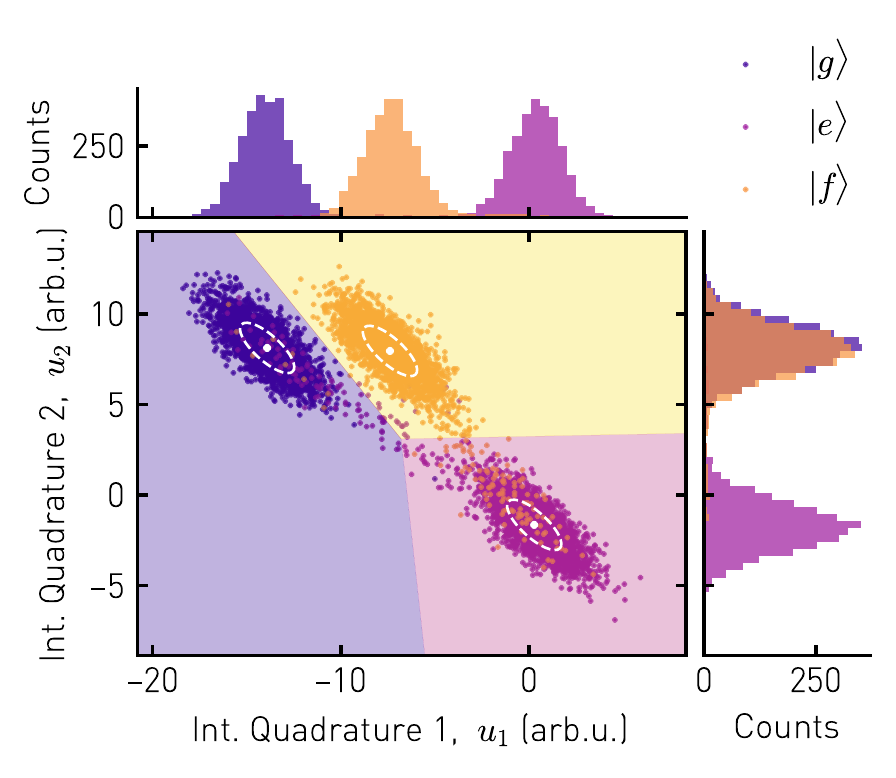}
	\caption{\textbf{Three-state single-shot readout.} Measured integrated readout signals $u_1$ and $u_2$ of the qubit prepared 3000 times in each of the three states $\ket{g}, \ket{e}$ and $\ket{f}$, after preselection. The histogram distributions of $u_1$ and $u_2$ are shown in the top and right panel, respectively. The mean value (dot) and 1$\sigma$ confidence ellipse (dashed) of each fitted Gaussian distribution is indicated in white. The colored regions indicate the assigned state in the integrated quadrature plane spanned by $u_1$ and $u_2$.}
	\label{fig:ssro}
\end{figure}

\subsection{Three-state single-shot readout} To dispersively read out the state of the transmon qubit, we apply a square pulse at frequency $\omega_\mathrm{r}$ of duration $\tau=1\,\mu$s with a Gaussian rising and falling edge ($\sigma=10\,$ns) to the readout resonator of the qubit~\cite{Heinsoo2018}. We multiply the digitized readout signal $s(t)$ in real-time on the UHFQA with two sets of complex-valued integration weights $w_1(t)$ and $w_2(t)$ to optimally distinguish between the first three states of the transmon qubit. We record the integrated readout signals
\begin{equation}
	u_i=\mathrm{Re}\left\{{\int_{0}^{1.2\,\mu\mathrm{s}}}s(t)w_i(t)\mathrm{d}t\right\}
\end{equation}
and evaluate the readout performance by preparing the qubit~3000 times in each of the three states $\ket{g}, \ket{e}$ and $\ket{f}$. Prior to the state preparation pulse, we prepend an additional readout, which allows us to condition each experimental repetition on the qubit being initially in the ground state $\ket{g}$. From this, we find a thermal population of 3.6$\,\%$. We analyze the preselected integrated readout signal histogram by fitting a trimodal Gaussian mixture model which allows us to assign qubit states to the integrated measurement results, see Fig.~\ref{fig:ssro}. From this experiment, we find the readout assignment probabilities $(p_{\ket{g}},p_{\ket{e}},p_{\ket{f}})=(99.7, 95.8, 96.0)\,\%$ resulting in an average three-state readout error of 2.8$\,\%$.

\bibliography{QuDevRefDB}

\begin{thebibliography}{33}%
\makeatletter
\providecommand \@ifxundefined [1]{%
 \@ifx{#1\undefined}
}%
\providecommand \@ifnum [1]{%
 \ifnum #1\expandafter \@firstoftwo
 \else \expandafter \@secondoftwo
 \fi
}%
\providecommand \@ifx [1]{%
 \ifx #1\expandafter \@firstoftwo
 \else \expandafter \@secondoftwo
 \fi
}%
\providecommand \natexlab [1]{#1}%
\providecommand \enquote  [1]{``#1''}%
\providecommand \bibnamefont  [1]{#1}%
\providecommand \bibfnamefont [1]{#1}%
\providecommand \citenamefont [1]{#1}%
\providecommand \href@noop [0]{\@secondoftwo}%
\providecommand \href [0]{\begingroup \@sanitize@url \@href}%
\providecommand \@href[1]{\@@startlink{#1}\@@href}%
\providecommand \@@href[1]{\endgroup#1\@@endlink}%
\providecommand \@sanitize@url [0]{\catcode `\\12\catcode `\$12\catcode
  `\&12\catcode `\#12\catcode `\^12\catcode `\_12\catcode `\%12\relax}%
\providecommand \@@startlink[1]{}%
\providecommand \@@endlink[0]{}%
\providecommand \url  [0]{\begingroup\@sanitize@url \@url }%
\providecommand \@url [1]{\endgroup\@href {#1}{\urlprefix }}%
\providecommand \urlprefix  [0]{URL }%
\providecommand \Eprint [0]{\href }%
\providecommand \doibase [0]{http://dx.doi.org/}%
\providecommand \selectlanguage [0]{\@gobble}%
\providecommand \bibinfo  [0]{\@secondoftwo}%
\providecommand \bibfield  [0]{\@secondoftwo}%
\providecommand \translation [1]{[#1]}%
\providecommand \BibitemOpen [0]{}%
\providecommand \bibitemStop [0]{}%
\providecommand \bibitemNoStop [0]{.\EOS\space}%
\providecommand \EOS [0]{\spacefactor3000\relax}%
\providecommand \BibitemShut  [1]{\csname bibitem#1\endcsname}%
\let\auto@bib@innerbib\@empty
\bibitem [{\citenamefont {Kelly}\ \emph {et~al.}(2015)\citenamefont {Kelly},
  \citenamefont {Barends}, \citenamefont {Fowler}, \citenamefont {Megrant},
  \citenamefont {Jeffrey}, \citenamefont {White}, \citenamefont {Sank},
  \citenamefont {Mutus}, \citenamefont {Campbell}, \citenamefont {Chen},
  \citenamefont {Chen}, \citenamefont {Chiaro}, \citenamefont {Dunsworth},
  \citenamefont {Hoi}, \citenamefont {Neill}, \citenamefont {O'Malley},
  \citenamefont {Quintana}, \citenamefont {Roushan}, \citenamefont
  {Vainsencher}, \citenamefont {Wenner}, \citenamefont {Cleland},\ and\
  \citenamefont {Martinis}}]{Kelly2015}%
  \BibitemOpen
  \bibfield  {author} {\bibinfo {author} {\bibfnamefont {J.}~\bibnamefont
  {Kelly}}, \bibinfo {author} {\bibfnamefont {R.}~\bibnamefont {Barends}},
  \bibinfo {author} {\bibfnamefont {A.~G.}\ \bibnamefont {Fowler}}, \bibinfo
  {author} {\bibfnamefont {A.}~\bibnamefont {Megrant}}, \bibinfo {author}
  {\bibfnamefont {E.}~\bibnamefont {Jeffrey}}, \bibinfo {author} {\bibfnamefont
  {T.~C.}\ \bibnamefont {White}}, \bibinfo {author} {\bibfnamefont
  {D.}~\bibnamefont {Sank}}, \bibinfo {author} {\bibfnamefont {J.~Y.}\
  \bibnamefont {Mutus}}, \bibinfo {author} {\bibfnamefont {B.}~\bibnamefont
  {Campbell}}, \bibinfo {author} {\bibfnamefont {Y.}~\bibnamefont {Chen}},
  \bibinfo {author} {\bibfnamefont {Z.}~\bibnamefont {Chen}}, \bibinfo {author}
  {\bibfnamefont {B.}~\bibnamefont {Chiaro}}, \bibinfo {author} {\bibfnamefont
  {A.}~\bibnamefont {Dunsworth}}, \bibinfo {author} {\bibfnamefont {I.-C.}\
  \bibnamefont {Hoi}}, \bibinfo {author} {\bibfnamefont {C.}~\bibnamefont
  {Neill}}, \bibinfo {author} {\bibfnamefont {P.~J.~J.}\ \bibnamefont
  {O'Malley}}, \bibinfo {author} {\bibfnamefont {C.}~\bibnamefont {Quintana}},
  \bibinfo {author} {\bibfnamefont {P.}~\bibnamefont {Roushan}}, \bibinfo
  {author} {\bibfnamefont {A.}~\bibnamefont {Vainsencher}}, \bibinfo {author}
  {\bibfnamefont {J.}~\bibnamefont {Wenner}}, \bibinfo {author} {\bibfnamefont
  {A.~N.}\ \bibnamefont {Cleland}}, \ and\ \bibinfo {author} {\bibfnamefont
  {J.~M.}\ \bibnamefont {Martinis}},\ }\href {\doibase doi:10.1038/nature14270}
  {\bibfield  {journal} {\bibinfo  {journal} {Nature}\ }\textbf {\bibinfo
  {volume} {519}},\ \bibinfo {pages} {66} (\bibinfo {year} {2015})}\BibitemShut
  {NoStop}%
\bibitem [{\citenamefont {Andersen}\ \emph {et~al.}(2020)\citenamefont
  {Andersen}, \citenamefont {Remm}, \citenamefont {Lazar}, \citenamefont
  {Krinner}, \citenamefont {Lacroix}, \citenamefont {Norris}, \citenamefont
  {Gabureac}, \citenamefont {Eichler},\ and\ \citenamefont
  {Wallraff}}]{Andersen2020b}%
  \BibitemOpen
  \bibfield  {author} {\bibinfo {author} {\bibfnamefont {C.~K.}\ \bibnamefont
  {Andersen}}, \bibinfo {author} {\bibfnamefont {A.}~\bibnamefont {Remm}},
  \bibinfo {author} {\bibfnamefont {S.}~\bibnamefont {Lazar}}, \bibinfo
  {author} {\bibfnamefont {S.}~\bibnamefont {Krinner}}, \bibinfo {author}
  {\bibfnamefont {N.}~\bibnamefont {Lacroix}}, \bibinfo {author} {\bibfnamefont
  {G.~J.}\ \bibnamefont {Norris}}, \bibinfo {author} {\bibfnamefont
  {M.}~\bibnamefont {Gabureac}}, \bibinfo {author} {\bibfnamefont
  {C.}~\bibnamefont {Eichler}}, \ and\ \bibinfo {author} {\bibfnamefont
  {A.}~\bibnamefont {Wallraff}},\ }\href
  {https://www.nature.com/articles/s41567-020-0920-y} {\bibfield  {journal}
  {\bibinfo  {journal} {Nature Physics}\ }\textbf {\bibinfo {volume} {16}},\
  \bibinfo {pages} {875} (\bibinfo {year} {2020})}\BibitemShut {NoStop}%
\bibitem [{\citenamefont {Chen}\ \emph {et~al.}(2021)\citenamefont {Chen},
  \citenamefont {Satzinger}, \citenamefont {Atalaya}, \citenamefont {Korotkov},
  \citenamefont {Dunsworth}, \citenamefont {Sank}, \citenamefont {Quintana},
  \citenamefont {McEwen}, \citenamefont {Barends}, \citenamefont {Klimov},
  \citenamefont {Hong}, \citenamefont {Jones}, \citenamefont {Petukhov},
  \citenamefont {Kafri}, \citenamefont {Demura}, \citenamefont {Burkett},
  \citenamefont {Gidney}, \citenamefont {Fowler}, \citenamefont {Putterman},
  \citenamefont {Aleiner}, \citenamefont {Arute}, \citenamefont {Arya},
  \citenamefont {Babbush}, \citenamefont {Bardin}, \citenamefont {Bengtsson},
  \citenamefont {Bourassa}, \citenamefont {Broughton}, \citenamefont {Buckley},
  \citenamefont {Buell}, \citenamefont {Bushnell}, \citenamefont {Chiaro},
  \citenamefont {Collins}, \citenamefont {Courtney}, \citenamefont {Derk},
  \citenamefont {Eppens}, \citenamefont {Erickson}, \citenamefont {Farhi},
  \citenamefont {Foxen}, \citenamefont {Giustina}, \citenamefont {Gross},
  \citenamefont {Harrigan}, \citenamefont {Harrington}, \citenamefont {Hilton},
  \citenamefont {Ho}, \citenamefont {Huang}, \citenamefont {Huggins},
  \citenamefont {Ioffe}, \citenamefont {Isakov}, \citenamefont {Jeffrey},
  \citenamefont {Jiang}, \citenamefont {Kechedzhi}, \citenamefont {Kim},
  \citenamefont {Kostritsa}, \citenamefont {Landhuis}, \citenamefont {Laptev},
  \citenamefont {Lucero}, \citenamefont {Martin}, \citenamefont {McClean},
  \citenamefont {McCourt}, \citenamefont {Mi}, \citenamefont {Miao},
  \citenamefont {Mohseni}, \citenamefont {Mruczkiewicz}, \citenamefont {Mutus},
  \citenamefont {Naaman}, \citenamefont {Neeley}, \citenamefont {Neill},
  \citenamefont {Newman}, \citenamefont {Niu}, \citenamefont {O'Brien},
  \citenamefont {Opremcak}, \citenamefont {Ostby}, \citenamefont {Pat\'o},
  \citenamefont {Redd}, \citenamefont {Roushan}, \citenamefont {Rubin},
  \citenamefont {Shvarts}, \citenamefont {Strain}, \citenamefont {Szalay},
  \citenamefont {Trevithick}, \citenamefont {Villalonga}, \citenamefont
  {White}, \citenamefont {Yao}, \citenamefont {Yeh}, \citenamefont {Zalcman},
  \citenamefont {Neven}, \citenamefont {Boixo}, \citenamefont {Smelyanskiy},
  \citenamefont {Chen}, \citenamefont {Megrant},\ and\ \citenamefont
  {Kelly}}]{Chen2021d}%
  \BibitemOpen
  \bibfield  {author} {\bibinfo {author} {\bibfnamefont {Z.}~\bibnamefont
  {Chen}}, \bibinfo {author} {\bibfnamefont {K.~J.}\ \bibnamefont {Satzinger}},
  \bibinfo {author} {\bibfnamefont {J.}~\bibnamefont {Atalaya}}, \bibinfo
  {author} {\bibfnamefont {A.~N.}\ \bibnamefont {Korotkov}}, \bibinfo {author}
  {\bibfnamefont {A.}~\bibnamefont {Dunsworth}}, \bibinfo {author}
  {\bibfnamefont {D.}~\bibnamefont {Sank}}, \bibinfo {author} {\bibfnamefont
  {C.}~\bibnamefont {Quintana}}, \bibinfo {author} {\bibfnamefont
  {M.}~\bibnamefont {McEwen}}, \bibinfo {author} {\bibfnamefont
  {R.}~\bibnamefont {Barends}}, \bibinfo {author} {\bibfnamefont {P.~V.}\
  \bibnamefont {Klimov}}, \bibinfo {author} {\bibfnamefont {S.}~\bibnamefont
  {Hong}}, \bibinfo {author} {\bibfnamefont {C.}~\bibnamefont {Jones}},
  \bibinfo {author} {\bibfnamefont {A.}~\bibnamefont {Petukhov}}, \bibinfo
  {author} {\bibfnamefont {D.}~\bibnamefont {Kafri}}, \bibinfo {author}
  {\bibfnamefont {S.}~\bibnamefont {Demura}}, \bibinfo {author} {\bibfnamefont
  {B.}~\bibnamefont {Burkett}}, \bibinfo {author} {\bibfnamefont
  {C.}~\bibnamefont {Gidney}}, \bibinfo {author} {\bibfnamefont {A.~G.}\
  \bibnamefont {Fowler}}, \bibinfo {author} {\bibfnamefont {H.}~\bibnamefont
  {Putterman}}, \bibinfo {author} {\bibfnamefont {I.}~\bibnamefont {Aleiner}},
  \bibinfo {author} {\bibfnamefont {F.}~\bibnamefont {Arute}}, \bibinfo
  {author} {\bibfnamefont {K.}~\bibnamefont {Arya}}, \bibinfo {author}
  {\bibfnamefont {R.}~\bibnamefont {Babbush}}, \bibinfo {author} {\bibfnamefont
  {J.~C.}\ \bibnamefont {Bardin}}, \bibinfo {author} {\bibfnamefont
  {A.}~\bibnamefont {Bengtsson}}, \bibinfo {author} {\bibfnamefont
  {A.}~\bibnamefont {Bourassa}}, \bibinfo {author} {\bibfnamefont
  {M.}~\bibnamefont {Broughton}}, \bibinfo {author} {\bibfnamefont {B.~B.}\
  \bibnamefont {Buckley}}, \bibinfo {author} {\bibfnamefont {D.~A.}\
  \bibnamefont {Buell}}, \bibinfo {author} {\bibfnamefont {N.}~\bibnamefont
  {Bushnell}}, \bibinfo {author} {\bibfnamefont {B.}~\bibnamefont {Chiaro}},
  \bibinfo {author} {\bibfnamefont {R.}~\bibnamefont {Collins}}, \bibinfo
  {author} {\bibfnamefont {W.}~\bibnamefont {Courtney}}, \bibinfo {author}
  {\bibfnamefont {A.~R.}\ \bibnamefont {Derk}}, \bibinfo {author}
  {\bibfnamefont {D.}~\bibnamefont {Eppens}}, \bibinfo {author} {\bibfnamefont
  {C.}~\bibnamefont {Erickson}}, \bibinfo {author} {\bibfnamefont
  {E.}~\bibnamefont {Farhi}}, \bibinfo {author} {\bibfnamefont
  {B.}~\bibnamefont {Foxen}}, \bibinfo {author} {\bibfnamefont
  {M.}~\bibnamefont {Giustina}}, \bibinfo {author} {\bibfnamefont {J.~A.}\
  \bibnamefont {Gross}}, \bibinfo {author} {\bibfnamefont {M.~P.}\ \bibnamefont
  {Harrigan}}, \bibinfo {author} {\bibfnamefont {S.~D.}\ \bibnamefont
  {Harrington}}, \bibinfo {author} {\bibfnamefont {J.}~\bibnamefont {Hilton}},
  \bibinfo {author} {\bibfnamefont {A.}~\bibnamefont {Ho}}, \bibinfo {author}
  {\bibfnamefont {T.}~\bibnamefont {Huang}}, \bibinfo {author} {\bibfnamefont
  {W.~J.}\ \bibnamefont {Huggins}}, \bibinfo {author} {\bibfnamefont {L.~B.}\
  \bibnamefont {Ioffe}}, \bibinfo {author} {\bibfnamefont {S.~V.}\ \bibnamefont
  {Isakov}}, \bibinfo {author} {\bibfnamefont {E.}~\bibnamefont {Jeffrey}},
  \bibinfo {author} {\bibfnamefont {Z.}~\bibnamefont {Jiang}}, \bibinfo
  {author} {\bibfnamefont {K.}~\bibnamefont {Kechedzhi}}, \bibinfo {author}
  {\bibfnamefont {S.}~\bibnamefont {Kim}}, \bibinfo {author} {\bibfnamefont
  {F.}~\bibnamefont {Kostritsa}}, \bibinfo {author} {\bibfnamefont
  {D.}~\bibnamefont {Landhuis}}, \bibinfo {author} {\bibfnamefont
  {P.}~\bibnamefont {Laptev}}, \bibinfo {author} {\bibfnamefont
  {E.}~\bibnamefont {Lucero}}, \bibinfo {author} {\bibfnamefont
  {O.}~\bibnamefont {Martin}}, \bibinfo {author} {\bibfnamefont {J.~R.}\
  \bibnamefont {McClean}}, \bibinfo {author} {\bibfnamefont {T.}~\bibnamefont
  {McCourt}}, \bibinfo {author} {\bibfnamefont {X.}~\bibnamefont {Mi}},
  \bibinfo {author} {\bibfnamefont {K.~C.}\ \bibnamefont {Miao}}, \bibinfo
  {author} {\bibfnamefont {M.}~\bibnamefont {Mohseni}}, \bibinfo {author}
  {\bibfnamefont {W.}~\bibnamefont {Mruczkiewicz}}, \bibinfo {author}
  {\bibfnamefont {J.}~\bibnamefont {Mutus}}, \bibinfo {author} {\bibfnamefont
  {O.}~\bibnamefont {Naaman}}, \bibinfo {author} {\bibfnamefont
  {M.}~\bibnamefont {Neeley}}, \bibinfo {author} {\bibfnamefont
  {C.}~\bibnamefont {Neill}}, \bibinfo {author} {\bibfnamefont
  {M.}~\bibnamefont {Newman}}, \bibinfo {author} {\bibfnamefont {M.~Y.}\
  \bibnamefont {Niu}}, \bibinfo {author} {\bibfnamefont {T.~E.}\ \bibnamefont
  {O'Brien}}, \bibinfo {author} {\bibfnamefont {A.}~\bibnamefont {Opremcak}},
  \bibinfo {author} {\bibfnamefont {E.}~\bibnamefont {Ostby}}, \bibinfo
  {author} {\bibfnamefont {B.}~\bibnamefont {Pat\'o}}, \bibinfo {author}
  {\bibfnamefont {N.}~\bibnamefont {Redd}}, \bibinfo {author} {\bibfnamefont
  {P.}~\bibnamefont {Roushan}}, \bibinfo {author} {\bibfnamefont {N.~C.}\
  \bibnamefont {Rubin}}, \bibinfo {author} {\bibfnamefont {V.}~\bibnamefont
  {Shvarts}}, \bibinfo {author} {\bibfnamefont {D.}~\bibnamefont {Strain}},
  \bibinfo {author} {\bibfnamefont {M.}~\bibnamefont {Szalay}}, \bibinfo
  {author} {\bibfnamefont {M.~D.}\ \bibnamefont {Trevithick}}, \bibinfo
  {author} {\bibfnamefont {B.}~\bibnamefont {Villalonga}}, \bibinfo {author}
  {\bibfnamefont {T.}~\bibnamefont {White}}, \bibinfo {author} {\bibfnamefont
  {Z.~J.}\ \bibnamefont {Yao}}, \bibinfo {author} {\bibfnamefont
  {P.}~\bibnamefont {Yeh}}, \bibinfo {author} {\bibfnamefont {A.}~\bibnamefont
  {Zalcman}}, \bibinfo {author} {\bibfnamefont {H.}~\bibnamefont {Neven}},
  \bibinfo {author} {\bibfnamefont {S.}~\bibnamefont {Boixo}}, \bibinfo
  {author} {\bibfnamefont {V.}~\bibnamefont {Smelyanskiy}}, \bibinfo {author}
  {\bibfnamefont {Y.}~\bibnamefont {Chen}}, \bibinfo {author} {\bibfnamefont
  {A.}~\bibnamefont {Megrant}}, \ and\ \bibinfo {author} {\bibfnamefont
  {J.}~\bibnamefont {Kelly}},\ }\href
  {https://doi.org/10.1038/s41586-021-03588-y} {\bibfield  {journal} {\bibinfo
  {journal} {Nature}\ }\textbf {\bibinfo {volume} {595}},\ \bibinfo {pages}
  {383} (\bibinfo {year} {2021})}\BibitemShut {NoStop}%
\bibitem [{\citenamefont {Krinner}\ \emph {et~al.}(2022)\citenamefont
  {Krinner}, \citenamefont {Lacroix}, \citenamefont {Remm}, \citenamefont
  {Paolo}, \citenamefont {Genois}, \citenamefont {Leroux}, \citenamefont
  {Hellings}, \citenamefont {Lazar}, \citenamefont {Swiadek}, \citenamefont
  {Herrmann}, \citenamefont {Norris}, \citenamefont {Andersen}, \citenamefont
  {Müller}, \citenamefont {Blais}, \citenamefont {Eichler},\ and\
  \citenamefont {Wallraff}}]{Krinner2021}%
  \BibitemOpen
  \bibfield  {author} {\bibinfo {author} {\bibfnamefont {S.}~\bibnamefont
  {Krinner}}, \bibinfo {author} {\bibfnamefont {N.}~\bibnamefont {Lacroix}},
  \bibinfo {author} {\bibfnamefont {A.}~\bibnamefont {Remm}}, \bibinfo {author}
  {\bibfnamefont {A.~D.}\ \bibnamefont {Paolo}}, \bibinfo {author}
  {\bibfnamefont {E.}~\bibnamefont {Genois}}, \bibinfo {author} {\bibfnamefont
  {C.}~\bibnamefont {Leroux}}, \bibinfo {author} {\bibfnamefont
  {C.}~\bibnamefont {Hellings}}, \bibinfo {author} {\bibfnamefont
  {S.}~\bibnamefont {Lazar}}, \bibinfo {author} {\bibfnamefont
  {F.}~\bibnamefont {Swiadek}}, \bibinfo {author} {\bibfnamefont
  {J.}~\bibnamefont {Herrmann}}, \bibinfo {author} {\bibfnamefont {G.~J.}\
  \bibnamefont {Norris}}, \bibinfo {author} {\bibfnamefont {C.~K.}\
  \bibnamefont {Andersen}}, \bibinfo {author} {\bibfnamefont {M.}~\bibnamefont
  {Müller}}, \bibinfo {author} {\bibfnamefont {A.}~\bibnamefont {Blais}},
  \bibinfo {author} {\bibfnamefont {C.}~\bibnamefont {Eichler}}, \ and\
  \bibinfo {author} {\bibfnamefont {A.}~\bibnamefont {Wallraff}},\ }\href
  {https://www.nature.com/articles/s41586-022-04566-8} {\bibfield  {journal}
  {\bibinfo  {journal} {Nature}\ }\textbf {\bibinfo {volume} {605}},\ \bibinfo
  {pages} {669} (\bibinfo {year} {2022})}\BibitemShut {NoStop}%
\bibitem [{\citenamefont {Preskill}(2018)}]{Preskill2018}%
  \BibitemOpen
  \bibfield  {author} {\bibinfo {author} {\bibfnamefont {J.}~\bibnamefont
  {Preskill}},\ }\href {\doibase 10.22331/q-2018-08-06-79} {\bibfield
  {journal} {\bibinfo  {journal} {{Quantum}}\ }\textbf {\bibinfo {volume}
  {2}},\ \bibinfo {pages} {79} (\bibinfo {year} {2018})}\BibitemShut {NoStop}%
\bibitem [{\citenamefont {Motzoi}\ \emph {et~al.}(2009)\citenamefont {Motzoi},
  \citenamefont {Gambetta}, \citenamefont {Rebentrost},\ and\ \citenamefont
  {Wilhelm}}]{Motzoi2009}%
  \BibitemOpen
  \bibfield  {author} {\bibinfo {author} {\bibfnamefont {F.}~\bibnamefont
  {Motzoi}}, \bibinfo {author} {\bibfnamefont {J.~M.}\ \bibnamefont
  {Gambetta}}, \bibinfo {author} {\bibfnamefont {P.}~\bibnamefont
  {Rebentrost}}, \ and\ \bibinfo {author} {\bibfnamefont {F.~K.}\ \bibnamefont
  {Wilhelm}},\ }\href {\doibase 10.1103/PhysRevLett.103.110501} {\bibfield
  {journal} {\bibinfo  {journal} {Phys. Rev. Lett.}\ }\textbf {\bibinfo
  {volume} {103}},\ \bibinfo {eid} {110501} (\bibinfo {year}
  {2009})}\BibitemShut {NoStop}%
\bibitem [{\citenamefont {Gambetta}\ \emph {et~al.}(2011)\citenamefont
  {Gambetta}, \citenamefont {Motzoi}, \citenamefont {Merkel},\ and\
  \citenamefont {Wilhelm}}]{Gambetta2011a}%
  \BibitemOpen
  \bibfield  {author} {\bibinfo {author} {\bibfnamefont {J.~M.}\ \bibnamefont
  {Gambetta}}, \bibinfo {author} {\bibfnamefont {F.}~\bibnamefont {Motzoi}},
  \bibinfo {author} {\bibfnamefont {S.~T.}\ \bibnamefont {Merkel}}, \ and\
  \bibinfo {author} {\bibfnamefont {F.~K.}\ \bibnamefont {Wilhelm}},\ }\href
  {\doibase 10.1103/PhysRevA.83.012308} {\bibfield  {journal} {\bibinfo
  {journal} {Phys. Rev. A}\ }\textbf {\bibinfo {volume} {83}},\ \bibinfo
  {pages} {012308} (\bibinfo {year} {2011})}\BibitemShut {NoStop}%
\bibitem [{\citenamefont {Blais}\ \emph {et~al.}(2004)\citenamefont {Blais},
  \citenamefont {Huang}, \citenamefont {Wallraff}, \citenamefont {Girvin},\
  and\ \citenamefont {Schoelkopf}}]{Blais2004}%
  \BibitemOpen
  \bibfield  {author} {\bibinfo {author} {\bibfnamefont {A.}~\bibnamefont
  {Blais}}, \bibinfo {author} {\bibfnamefont {R.-S.}\ \bibnamefont {Huang}},
  \bibinfo {author} {\bibfnamefont {A.}~\bibnamefont {Wallraff}}, \bibinfo
  {author} {\bibfnamefont {S.~M.}\ \bibnamefont {Girvin}}, \ and\ \bibinfo
  {author} {\bibfnamefont {R.~J.}\ \bibnamefont {Schoelkopf}},\ }\href
  {\doibase 10.1103/PhysRevA.69.062320} {\bibfield  {journal} {\bibinfo
  {journal} {Phys. Rev. A}\ }\textbf {\bibinfo {volume} {69}},\ \bibinfo
  {pages} {062320} (\bibinfo {year} {2004})}\BibitemShut {NoStop}%
\bibitem [{\citenamefont {Wallraff}\ \emph {et~al.}(2005)\citenamefont
  {Wallraff}, \citenamefont {Schuster}, \citenamefont {Blais}, \citenamefont
  {Frunzio}, \citenamefont {Majer}, \citenamefont {Devoret}, \citenamefont
  {Girvin},\ and\ \citenamefont {Schoelkopf}}]{Wallraff2005}%
  \BibitemOpen
  \bibfield  {author} {\bibinfo {author} {\bibfnamefont {A.}~\bibnamefont
  {Wallraff}}, \bibinfo {author} {\bibfnamefont {D.~I.}\ \bibnamefont
  {Schuster}}, \bibinfo {author} {\bibfnamefont {A.}~\bibnamefont {Blais}},
  \bibinfo {author} {\bibfnamefont {L.}~\bibnamefont {Frunzio}}, \bibinfo
  {author} {\bibfnamefont {J.}~\bibnamefont {Majer}}, \bibinfo {author}
  {\bibfnamefont {M.~H.}\ \bibnamefont {Devoret}}, \bibinfo {author}
  {\bibfnamefont {S.~M.}\ \bibnamefont {Girvin}}, \ and\ \bibinfo {author}
  {\bibfnamefont {R.~J.}\ \bibnamefont {Schoelkopf}},\ }\href {\doibase
  10.1103/PhysRevLett.95.060501} {\bibfield  {journal} {\bibinfo  {journal}
  {Phys. Rev. Lett.}\ }\textbf {\bibinfo {volume} {95}},\ \bibinfo {pages}
  {060501} (\bibinfo {year} {2005})}\BibitemShut {NoStop}%
\bibitem [{\citenamefont {Walter}\ \emph {et~al.}(2017)\citenamefont {Walter},
  \citenamefont {Kurpiers}, \citenamefont {Gasparinetti}, \citenamefont
  {Magnard}, \citenamefont {Poto\v{c}nik}, \citenamefont {Salath\'e},
  \citenamefont {Pechal}, \citenamefont {Mondal}, \citenamefont {Oppliger},
  \citenamefont {Eichler},\ and\ \citenamefont {Wallraff}}]{Walter2017}%
  \BibitemOpen
  \bibfield  {author} {\bibinfo {author} {\bibfnamefont {T.}~\bibnamefont
  {Walter}}, \bibinfo {author} {\bibfnamefont {P.}~\bibnamefont {Kurpiers}},
  \bibinfo {author} {\bibfnamefont {S.}~\bibnamefont {Gasparinetti}}, \bibinfo
  {author} {\bibfnamefont {P.}~\bibnamefont {Magnard}}, \bibinfo {author}
  {\bibfnamefont {A.}~\bibnamefont {Poto\v{c}nik}}, \bibinfo {author}
  {\bibfnamefont {Y.}~\bibnamefont {Salath\'e}}, \bibinfo {author}
  {\bibfnamefont {M.}~\bibnamefont {Pechal}}, \bibinfo {author} {\bibfnamefont
  {M.}~\bibnamefont {Mondal}}, \bibinfo {author} {\bibfnamefont
  {M.}~\bibnamefont {Oppliger}}, \bibinfo {author} {\bibfnamefont
  {C.}~\bibnamefont {Eichler}}, \ and\ \bibinfo {author} {\bibfnamefont
  {A.}~\bibnamefont {Wallraff}},\ }\href {\doibase
  10.1103/PhysRevApplied.7.054020} {\bibfield  {journal} {\bibinfo  {journal}
  {Phys. Rev. Appl.}\ }\textbf {\bibinfo {volume} {7}},\ \bibinfo {pages}
  {054020} (\bibinfo {year} {2017})}\BibitemShut {NoStop}%
\bibitem [{\citenamefont {Chow}\ \emph {et~al.}(2011)\citenamefont {Chow},
  \citenamefont {C\'orcoles}, \citenamefont {Gambetta}, \citenamefont
  {Rigetti}, \citenamefont {Johnson}, \citenamefont {Smolin}, \citenamefont
  {Rozen}, \citenamefont {Keefe}, \citenamefont {Rothwell}, \citenamefont
  {Ketchen},\ and\ \citenamefont {Steffen}}]{Chow2011}%
  \BibitemOpen
  \bibfield  {author} {\bibinfo {author} {\bibfnamefont {J.~M.}\ \bibnamefont
  {Chow}}, \bibinfo {author} {\bibfnamefont {A.~D.}\ \bibnamefont
  {C\'orcoles}}, \bibinfo {author} {\bibfnamefont {J.~M.}\ \bibnamefont
  {Gambetta}}, \bibinfo {author} {\bibfnamefont {C.}~\bibnamefont {Rigetti}},
  \bibinfo {author} {\bibfnamefont {B.~R.}\ \bibnamefont {Johnson}}, \bibinfo
  {author} {\bibfnamefont {J.~A.}\ \bibnamefont {Smolin}}, \bibinfo {author}
  {\bibfnamefont {J.~R.}\ \bibnamefont {Rozen}}, \bibinfo {author}
  {\bibfnamefont {G.~A.}\ \bibnamefont {Keefe}}, \bibinfo {author}
  {\bibfnamefont {M.~B.}\ \bibnamefont {Rothwell}}, \bibinfo {author}
  {\bibfnamefont {M.~B.}\ \bibnamefont {Ketchen}}, \ and\ \bibinfo {author}
  {\bibfnamefont {M.}~\bibnamefont {Steffen}},\ }\href {\doibase
  10.1103/PhysRevLett.107.080502} {\bibfield  {journal} {\bibinfo  {journal}
  {Phys. Rev. Lett.}\ }\textbf {\bibinfo {volume} {107}},\ \bibinfo {pages}
  {080502} (\bibinfo {year} {2011})}\BibitemShut {NoStop}%
\bibitem [{\citenamefont {McKay}\ \emph {et~al.}(2017)\citenamefont {McKay},
  \citenamefont {Wood}, \citenamefont {Sheldon}, \citenamefont {Chow},\ and\
  \citenamefont {Gambetta}}]{McKay2017}%
  \BibitemOpen
  \bibfield  {author} {\bibinfo {author} {\bibfnamefont {D.~C.}\ \bibnamefont
  {McKay}}, \bibinfo {author} {\bibfnamefont {C.~J.}\ \bibnamefont {Wood}},
  \bibinfo {author} {\bibfnamefont {S.}~\bibnamefont {Sheldon}}, \bibinfo
  {author} {\bibfnamefont {J.~M.}\ \bibnamefont {Chow}}, \ and\ \bibinfo
  {author} {\bibfnamefont {J.~M.}\ \bibnamefont {Gambetta}},\ }\href {\doibase
  10.1103/PhysRevA.96.022330} {\bibfield  {journal} {\bibinfo  {journal} {Phys.
  Rev. A}\ }\textbf {\bibinfo {volume} {96}},\ \bibinfo {pages} {022330}
  (\bibinfo {year} {2017})}\BibitemShut {NoStop}%
\bibitem [{\citenamefont {Malinowski}\ \emph {et~al.}(2017)\citenamefont
  {Malinowski}, \citenamefont {Martins}, \citenamefont {Nissen}, \citenamefont
  {Fallahi}, \citenamefont {Gardner}, \citenamefont {Manfra}, \citenamefont
  {Marcus},\ and\ \citenamefont {Kuemmeth}}]{Malinowski2017a}%
  \BibitemOpen
  \bibfield  {author} {\bibinfo {author} {\bibfnamefont {F.~K.}\ \bibnamefont
  {Malinowski}}, \bibinfo {author} {\bibfnamefont {F.}~\bibnamefont {Martins}},
  \bibinfo {author} {\bibfnamefont {P.~D.}\ \bibnamefont {Nissen}}, \bibinfo
  {author} {\bibfnamefont {S.}~\bibnamefont {Fallahi}}, \bibinfo {author}
  {\bibfnamefont {G.~C.}\ \bibnamefont {Gardner}}, \bibinfo {author}
  {\bibfnamefont {M.~J.}\ \bibnamefont {Manfra}}, \bibinfo {author}
  {\bibfnamefont {C.~M.}\ \bibnamefont {Marcus}}, \ and\ \bibinfo {author}
  {\bibfnamefont {F.}~\bibnamefont {Kuemmeth}},\ }\href {\doibase
  10.1103/PhysRevB.96.045443} {\bibfield  {journal} {\bibinfo  {journal} {Phys.
  Rev. B}\ }\textbf {\bibinfo {volume} {96}},\ \bibinfo {pages} {045443}
  (\bibinfo {year} {2017})}\BibitemShut {NoStop}%
\bibitem [{\citenamefont {Barthel}\ \emph {et~al.}(2010)\citenamefont
  {Barthel}, \citenamefont {Kj\ae{}rgaard}, \citenamefont {Medford},
  \citenamefont {Stopa}, \citenamefont {Marcus}, \citenamefont {Hanson},\ and\
  \citenamefont {Gossard}}]{Barthel2010}%
  \BibitemOpen
  \bibfield  {author} {\bibinfo {author} {\bibfnamefont {C.}~\bibnamefont
  {Barthel}}, \bibinfo {author} {\bibfnamefont {M.}~\bibnamefont
  {Kj\ae{}rgaard}}, \bibinfo {author} {\bibfnamefont {J.}~\bibnamefont
  {Medford}}, \bibinfo {author} {\bibfnamefont {M.}~\bibnamefont {Stopa}},
  \bibinfo {author} {\bibfnamefont {C.~M.}\ \bibnamefont {Marcus}}, \bibinfo
  {author} {\bibfnamefont {M.~P.}\ \bibnamefont {Hanson}}, \ and\ \bibinfo
  {author} {\bibfnamefont {A.~C.}\ \bibnamefont {Gossard}},\ }\href {\doibase
  10.1103/PhysRevB.81.161308} {\bibfield  {journal} {\bibinfo  {journal} {Phys.
  Rev. B}\ }\textbf {\bibinfo {volume} {81}},\ \bibinfo {pages} {161308}
  (\bibinfo {year} {2010})}\BibitemShut {NoStop}%
\bibitem [{\citenamefont {Harty}\ \emph {et~al.}(2016)\citenamefont {Harty},
  \citenamefont {Sepiol}, \citenamefont {Allcock}, \citenamefont {Ballance},
  \citenamefont {Tarlton},\ and\ \citenamefont {Lucas}}]{Harty2016}%
  \BibitemOpen
  \bibfield  {author} {\bibinfo {author} {\bibfnamefont {T.~P.}\ \bibnamefont
  {Harty}}, \bibinfo {author} {\bibfnamefont {M.~A.}\ \bibnamefont {Sepiol}},
  \bibinfo {author} {\bibfnamefont {D.~T.~C.}\ \bibnamefont {Allcock}},
  \bibinfo {author} {\bibfnamefont {C.~J.}\ \bibnamefont {Ballance}}, \bibinfo
  {author} {\bibfnamefont {J.~E.}\ \bibnamefont {Tarlton}}, \ and\ \bibinfo
  {author} {\bibfnamefont {D.~M.}\ \bibnamefont {Lucas}},\ }\href@noop {}
  {\bibfield  {journal} {\bibinfo  {journal} {Phys. Rev. Lett.}\ }\textbf
  {\bibinfo {volume} {117}},\ \bibinfo {pages} {140501} (\bibinfo {year}
  {2016})}\BibitemShut {NoStop}%
\bibitem [{\citenamefont {Ospelkaus}\ \emph {et~al.}(2011)\citenamefont
  {Ospelkaus}, \citenamefont {Warring}, \citenamefont {Colombe}, \citenamefont
  {Brown}, \citenamefont {Amini}, \citenamefont {Leibfried},\ and\
  \citenamefont {Wineland}}]{Ospelkaus2011}%
  \BibitemOpen
  \bibfield  {author} {\bibinfo {author} {\bibfnamefont {C.}~\bibnamefont
  {Ospelkaus}}, \bibinfo {author} {\bibfnamefont {U.}~\bibnamefont {Warring}},
  \bibinfo {author} {\bibfnamefont {Y.}~\bibnamefont {Colombe}}, \bibinfo
  {author} {\bibfnamefont {K.~R.}\ \bibnamefont {Brown}}, \bibinfo {author}
  {\bibfnamefont {J.~M.}\ \bibnamefont {Amini}}, \bibinfo {author}
  {\bibfnamefont {D.}~\bibnamefont {Leibfried}}, \ and\ \bibinfo {author}
  {\bibfnamefont {D.~J.}\ \bibnamefont {Wineland}},\ }\href {\doibase
  10.1038/nature10290} {\bibfield  {journal} {\bibinfo  {journal} {Nature}\
  }\textbf {\bibinfo {volume} {476}},\ \bibinfo {pages} {181} (\bibinfo {year}
  {2011})}\BibitemShut {NoStop}%
\bibitem [{\citenamefont {Jolin}\ \emph {et~al.}(2020)\citenamefont {Jolin},
  \citenamefont {Borgani}, \citenamefont {Tholén}, \citenamefont
  {Forchheimer},\ and\ \citenamefont {Haviland}}]{Jolin2020}%
  \BibitemOpen
  \bibfield  {author} {\bibinfo {author} {\bibfnamefont {S.~W.}\ \bibnamefont
  {Jolin}}, \bibinfo {author} {\bibfnamefont {R.}~\bibnamefont {Borgani}},
  \bibinfo {author} {\bibfnamefont {M.~O.}\ \bibnamefont {Tholén}}, \bibinfo
  {author} {\bibfnamefont {D.}~\bibnamefont {Forchheimer}}, \ and\ \bibinfo
  {author} {\bibfnamefont {D.~B.}\ \bibnamefont {Haviland}},\ }\href {\doibase
  10.1063/5.0025836} {\bibfield  {journal} {\bibinfo  {journal} {Review of
  Scientific Instruments}\ }\textbf {\bibinfo {volume} {91}},\ \bibinfo {pages}
  {124707} (\bibinfo {year} {2020})}\BibitemShut {NoStop}%
\bibitem [{\citenamefont {Raftery}\ \emph {et~al.}(2017)\citenamefont
  {Raftery}, \citenamefont {Vrajitoarea}, \citenamefont {Zhang}, \citenamefont
  {Leng}, \citenamefont {Srinivasan},\ and\ \citenamefont
  {Houck}}]{Raftery2017}%
  \BibitemOpen
  \bibfield  {author} {\bibinfo {author} {\bibfnamefont {J.}~\bibnamefont
  {Raftery}}, \bibinfo {author} {\bibfnamefont {A.}~\bibnamefont
  {Vrajitoarea}}, \bibinfo {author} {\bibfnamefont {G.}~\bibnamefont {Zhang}},
  \bibinfo {author} {\bibfnamefont {Z.}~\bibnamefont {Leng}}, \bibinfo {author}
  {\bibfnamefont {S.~J.}\ \bibnamefont {Srinivasan}}, \ and\ \bibinfo {author}
  {\bibfnamefont {A.~A.}\ \bibnamefont {Houck}},\ }\href
  {https://arxiv.org/abs/1703.00942} {\bibfield  {journal} {\bibinfo  {journal}
  {arXiv:1703.00942}\ } (\bibinfo {year} {2017})}\BibitemShut {NoStop}%
\bibitem [{\citenamefont {Kalfus}\ \emph {et~al.}(2020)\citenamefont {Kalfus},
  \citenamefont {Lee}, \citenamefont {Ribeill}, \citenamefont {Fallek},
  \citenamefont {Wagner}, \citenamefont {Donovan}, \citenamefont {Riste},\ and\
  \citenamefont {Ohki}}]{Kalfus2020}%
  \BibitemOpen
  \bibfield  {author} {\bibinfo {author} {\bibfnamefont {W.~D.}\ \bibnamefont
  {Kalfus}}, \bibinfo {author} {\bibfnamefont {D.~F.}\ \bibnamefont {Lee}},
  \bibinfo {author} {\bibfnamefont {G.~J.}\ \bibnamefont {Ribeill}}, \bibinfo
  {author} {\bibfnamefont {S.~D.}\ \bibnamefont {Fallek}}, \bibinfo {author}
  {\bibfnamefont {A.}~\bibnamefont {Wagner}}, \bibinfo {author} {\bibfnamefont
  {B.}~\bibnamefont {Donovan}}, \bibinfo {author} {\bibfnamefont
  {D.}~\bibnamefont {Riste}}, \ and\ \bibinfo {author} {\bibfnamefont {T.~A.}\
  \bibnamefont {Ohki}},\ }\href {https://ieeexplore.ieee.org/document/9286506}
  {\bibfield  {journal} {\bibinfo  {journal} {IEEE Trans. Quantum Eng.}\
  }\textbf {\bibinfo {volume} {1}},\ \bibinfo {pages} {1} (\bibinfo {year}
  {2020})}\BibitemShut {NoStop}%
\bibitem [{\citenamefont {Xu}\ \emph {et~al.}(2021)\citenamefont {Xu},
  \citenamefont {Huang}, \citenamefont {Santiago},\ and\ \citenamefont
  {Siddiqi}}]{Xu2021d}%
  \BibitemOpen
  \bibfield  {author} {\bibinfo {author} {\bibfnamefont {Y.}~\bibnamefont
  {Xu}}, \bibinfo {author} {\bibfnamefont {G.}~\bibnamefont {Huang}}, \bibinfo
  {author} {\bibfnamefont {D.~I.}\ \bibnamefont {Santiago}}, \ and\ \bibinfo
  {author} {\bibfnamefont {I.}~\bibnamefont {Siddiqi}},\ }\href
  {https://aip.scitation.org/doi/10.1063/5.0055906} {\bibfield  {journal}
  {\bibinfo  {journal} {Rev. Sci. Instrum.}\ }\textbf {\bibinfo {volume}
  {92}},\ \bibinfo {pages} {075108} (\bibinfo {year} {2021})}\BibitemShut
  {NoStop}%
\bibitem [{\citenamefont {Chen}\ \emph {et~al.}(2016)\citenamefont {Chen},
  \citenamefont {Kelly}, \citenamefont {Quintana}, \citenamefont {Barends},
  \citenamefont {Campbell}, \citenamefont {Chen}, \citenamefont {Chiaro},
  \citenamefont {Dunsworth}, \citenamefont {Fowler}, \citenamefont {Lucero},
  \citenamefont {Jeffrey}, \citenamefont {Megrant}, \citenamefont {Mutus},
  \citenamefont {Neeley}, \citenamefont {Neill}, \citenamefont {O'Malley},
  \citenamefont {Roushan}, \citenamefont {Sank}, \citenamefont {Vainsencher},
  \citenamefont {Wenner}, \citenamefont {White}, \citenamefont {Korotkov},\
  and\ \citenamefont {Martinis}}]{Chen2016}%
  \BibitemOpen
  \bibfield  {author} {\bibinfo {author} {\bibfnamefont {Z.}~\bibnamefont
  {Chen}}, \bibinfo {author} {\bibfnamefont {J.}~\bibnamefont {Kelly}},
  \bibinfo {author} {\bibfnamefont {C.}~\bibnamefont {Quintana}}, \bibinfo
  {author} {\bibfnamefont {R.}~\bibnamefont {Barends}}, \bibinfo {author}
  {\bibfnamefont {B.}~\bibnamefont {Campbell}}, \bibinfo {author}
  {\bibfnamefont {Y.}~\bibnamefont {Chen}}, \bibinfo {author} {\bibfnamefont
  {B.}~\bibnamefont {Chiaro}}, \bibinfo {author} {\bibfnamefont
  {A.}~\bibnamefont {Dunsworth}}, \bibinfo {author} {\bibfnamefont {A.~G.}\
  \bibnamefont {Fowler}}, \bibinfo {author} {\bibfnamefont {E.}~\bibnamefont
  {Lucero}}, \bibinfo {author} {\bibfnamefont {E.}~\bibnamefont {Jeffrey}},
  \bibinfo {author} {\bibfnamefont {A.}~\bibnamefont {Megrant}}, \bibinfo
  {author} {\bibfnamefont {J.}~\bibnamefont {Mutus}}, \bibinfo {author}
  {\bibfnamefont {M.}~\bibnamefont {Neeley}}, \bibinfo {author} {\bibfnamefont
  {C.}~\bibnamefont {Neill}}, \bibinfo {author} {\bibfnamefont {P.~J.~J.}\
  \bibnamefont {O'Malley}}, \bibinfo {author} {\bibfnamefont {P.}~\bibnamefont
  {Roushan}}, \bibinfo {author} {\bibfnamefont {D.}~\bibnamefont {Sank}},
  \bibinfo {author} {\bibfnamefont {A.}~\bibnamefont {Vainsencher}}, \bibinfo
  {author} {\bibfnamefont {J.}~\bibnamefont {Wenner}}, \bibinfo {author}
  {\bibfnamefont {T.~C.}\ \bibnamefont {White}}, \bibinfo {author}
  {\bibfnamefont {A.~N.}\ \bibnamefont {Korotkov}}, \ and\ \bibinfo {author}
  {\bibfnamefont {J.~M.}\ \bibnamefont {Martinis}},\ }\href {\doibase
  10.1103/PhysRevLett.116.020501} {\bibfield  {journal} {\bibinfo  {journal}
  {Phys. Rev. Lett.}\ }\textbf {\bibinfo {volume} {116}},\ \bibinfo {pages}
  {020501} (\bibinfo {year} {2016})}\BibitemShut {NoStop}%
\bibitem [{\citenamefont {Hartley}(1928)}]{Hartley1928}%
  \BibitemOpen
  \bibfield  {author} {\bibinfo {author} {\bibfnamefont {R.~V.~L.}\
  \bibnamefont {Hartley}},\ }\href
  {https://www.ee.iitm.ac.in/~ani/2011/ee6240/pdf/Hartley_Patent_MODULATION_SYSTEM.pdf}
  {\enquote {\bibinfo {title} {Modulation system},}\ } (\bibinfo {year}
  {1928}),\ \bibinfo {note} {{US} Patent 1,666,206}\BibitemShut {NoStop}%
\bibitem [{\citenamefont {Baldwin}\ and\ \citenamefont
  {Dubbert}(2002)}]{Sandia2002}%
  \BibitemOpen
  \bibfield  {author} {\bibinfo {author} {\bibfnamefont {J.~G.}\ \bibnamefont
  {Baldwin}}\ and\ \bibinfo {author} {\bibfnamefont {D.~F.}\ \bibnamefont
  {Dubbert}},\ }\href {https://www.osti.gov/servlets/purl/800958/} {\enquote
  {\bibinfo {title} {Quadrature mixer lo leakage suppression through quadrature
  dc bias},}\ } (\bibinfo {year} {2002}),\ \bibinfo {note} {{S}andia
  Report}\BibitemShut {NoStop}%
\bibitem [{\citenamefont {Davis}(2010)}]{Alan2010}%
  \BibitemOpen
  \bibfield  {author} {\bibinfo {author} {\bibfnamefont {W.~A.}\ \bibnamefont
  {Davis}},\ }\enquote {\bibinfo {title} {{RF} mixers},}\ in\ \href {\doibase
  10.1002/9780470768020.ch11} {\emph {\bibinfo {booktitle} {Radio Frequency
  Circuit Design}}}\ (\bibinfo {year} {2010})\ pp.\ \bibinfo {pages}
  {289--319}\BibitemShut {NoStop}%
\bibitem [{\citenamefont {Bensky}(2019)}]{Alan2019}%
  \BibitemOpen
  \bibfield  {author} {\bibinfo {author} {\bibfnamefont {A.}~\bibnamefont
  {Bensky}},\ }\enquote {\bibinfo {title} {Radio system design},}\ in\ \href
  {\doibase https://doi.org/10.1016/B978-0-12-815405-2.00007-5} {\emph
  {\bibinfo {booktitle} {Short-range Wireless Communication (Third
  Edition)}}},\ \bibinfo {editor} {edited by\ \bibinfo {editor} {\bibfnamefont
  {A.}~\bibnamefont {Bensky}}}\ (\bibinfo {year} {2019})\ pp.\ \bibinfo {pages}
  {163--198}\BibitemShut {NoStop}%
\bibitem [{\citenamefont {Marki}\ and\ \citenamefont
  {Marki}(2010)}]{MarkiMixer}%
  \BibitemOpen
  \bibfield  {author} {\bibinfo {author} {\bibfnamefont {F.}~\bibnamefont
  {Marki}}\ and\ \bibinfo {author} {\bibfnamefont {C.}~\bibnamefont {Marki}},\
  }\href
  {https://www.markimicrowave.com/Assets/appnotes/mixer_basics_primer.pdf}
  {\enquote {\bibinfo {title} {Microwave mixer basics primer},}\ } (\bibinfo
  {year} {2010}),\ \bibinfo {note} {{A}pplication Note, Marki
  Microwave}\BibitemShut {NoStop}%
\bibitem [{\citenamefont {Ball}\ \emph {et~al.}(2016)\citenamefont {Ball},
  \citenamefont {Oliver},\ and\ \citenamefont {Biercuk}}]{Ball2016b}%
  \BibitemOpen
  \bibfield  {author} {\bibinfo {author} {\bibfnamefont {H.}~\bibnamefont
  {Ball}}, \bibinfo {author} {\bibfnamefont {W.~D.}\ \bibnamefont {Oliver}}, \
  and\ \bibinfo {author} {\bibfnamefont {M.~J.}\ \bibnamefont {Biercuk}},\
  }\href {\doibase 10.1103/PhysRevApplied.6.064009} {\bibfield  {journal}
  {\bibinfo  {journal} {npj Quantum Information}\ }\textbf {\bibinfo {volume}
  {2}},\ \bibinfo {pages} {16033} (\bibinfo {year} {2016})}\BibitemShut
  {NoStop}%
\bibitem [{\citenamefont {Emerson}\ \emph {et~al.}(2005)\citenamefont
  {Emerson}, \citenamefont {Alicki},\ and\ \citenamefont
  {Zyczkowski}}]{Emerson2005}%
  \BibitemOpen
  \bibfield  {author} {\bibinfo {author} {\bibfnamefont {J.}~\bibnamefont
  {Emerson}}, \bibinfo {author} {\bibfnamefont {R.}~\bibnamefont {Alicki}}, \
  and\ \bibinfo {author} {\bibfnamefont {K.}~\bibnamefont {Zyczkowski}},\
  }\href@noop {} {\bibfield  {journal} {\bibinfo  {journal} {J. Opt. B: Quantum
  Semiclass. Opt.}\ }\textbf {\bibinfo {volume} {7}},\ \bibinfo {pages} {S347}
  (\bibinfo {year} {2005})}\BibitemShut {NoStop}%
\bibitem [{\citenamefont {Magesan}\ \emph {et~al.}(2011)\citenamefont
  {Magesan}, \citenamefont {Gambetta},\ and\ \citenamefont
  {Emerson}}]{Magesan2011}%
  \BibitemOpen
  \bibfield  {author} {\bibinfo {author} {\bibfnamefont {E.}~\bibnamefont
  {Magesan}}, \bibinfo {author} {\bibfnamefont {J.~M.}\ \bibnamefont
  {Gambetta}}, \ and\ \bibinfo {author} {\bibfnamefont {J.}~\bibnamefont
  {Emerson}},\ }\href {\doibase 10.1103/PhysRevLett.106.180504} {\bibfield
  {journal} {\bibinfo  {journal} {Phys. Rev. Lett.}\ }\textbf {\bibinfo
  {volume} {106}},\ \bibinfo {pages} {180504} (\bibinfo {year}
  {2011})}\BibitemShut {NoStop}%
\bibitem [{\citenamefont {Asaad}\ \emph {et~al.}(2016)\citenamefont {Asaad},
  \citenamefont {Dickel}, \citenamefont {Langford}, \citenamefont {Poletto},
  \citenamefont {Bruno}, \citenamefont {Rol}, \citenamefont {Deurloo},\ and\
  \citenamefont {DiCarlo}}]{Asaad2016}%
  \BibitemOpen
  \bibfield  {author} {\bibinfo {author} {\bibfnamefont {S.}~\bibnamefont
  {Asaad}}, \bibinfo {author} {\bibfnamefont {C.}~\bibnamefont {Dickel}},
  \bibinfo {author} {\bibfnamefont {N.~K.}\ \bibnamefont {Langford}}, \bibinfo
  {author} {\bibfnamefont {S.}~\bibnamefont {Poletto}}, \bibinfo {author}
  {\bibfnamefont {A.}~\bibnamefont {Bruno}}, \bibinfo {author} {\bibfnamefont
  {M.~A.}\ \bibnamefont {Rol}}, \bibinfo {author} {\bibfnamefont
  {D.}~\bibnamefont {Deurloo}}, \ and\ \bibinfo {author} {\bibfnamefont
  {L.}~\bibnamefont {DiCarlo}},\ }\href
  {http://dx.doi.org/10.1038/npjqi.2016.29} {\bibfield  {journal} {\bibinfo
  {journal} {npj Quantum Information}\ }\textbf {\bibinfo {volume} {2}},\
  \bibinfo {pages} {16029} (\bibinfo {year} {2016})}\BibitemShut {NoStop}%
\bibitem [{\citenamefont {Macklin}\ \emph {et~al.}(2015)\citenamefont
  {Macklin}, \citenamefont {O'Brien}, \citenamefont {Hover}, \citenamefont
  {Schwartz}, \citenamefont {Bolkhovsky}, \citenamefont {Zhang}, \citenamefont
  {Oliver},\ and\ \citenamefont {Siddiqi}}]{Macklin2015}%
  \BibitemOpen
  \bibfield  {author} {\bibinfo {author} {\bibfnamefont {C.}~\bibnamefont
  {Macklin}}, \bibinfo {author} {\bibfnamefont {K.}~\bibnamefont {O'Brien}},
  \bibinfo {author} {\bibfnamefont {D.}~\bibnamefont {Hover}}, \bibinfo
  {author} {\bibfnamefont {M.~E.}\ \bibnamefont {Schwartz}}, \bibinfo {author}
  {\bibfnamefont {V.}~\bibnamefont {Bolkhovsky}}, \bibinfo {author}
  {\bibfnamefont {X.}~\bibnamefont {Zhang}}, \bibinfo {author} {\bibfnamefont
  {W.~D.}\ \bibnamefont {Oliver}}, \ and\ \bibinfo {author} {\bibfnamefont
  {I.}~\bibnamefont {Siddiqi}},\ }\href {\doibase 10.1126/science.aaa8525}
  {\bibfield  {journal} {\bibinfo  {journal} {Science}\ }\textbf {\bibinfo
  {volume} {350}},\ \bibinfo {pages} {307} (\bibinfo {year}
  {2015})}\BibitemShut {NoStop}%
\bibitem [{\citenamefont {Vion}\ \emph {et~al.}(2002)\citenamefont {Vion},
  \citenamefont {Aassime}, \citenamefont {Cottet}, \citenamefont {Joyez},
  \citenamefont {Pothier}, \citenamefont {Urbina}, \citenamefont {Esteve},\
  and\ \citenamefont {Devoret}}]{Vion2002}%
  \BibitemOpen
  \bibfield  {author} {\bibinfo {author} {\bibfnamefont {D.}~\bibnamefont
  {Vion}}, \bibinfo {author} {\bibfnamefont {A.}~\bibnamefont {Aassime}},
  \bibinfo {author} {\bibfnamefont {A.}~\bibnamefont {Cottet}}, \bibinfo
  {author} {\bibfnamefont {P.}~\bibnamefont {Joyez}}, \bibinfo {author}
  {\bibfnamefont {H.}~\bibnamefont {Pothier}}, \bibinfo {author} {\bibfnamefont
  {C.}~\bibnamefont {Urbina}}, \bibinfo {author} {\bibfnamefont
  {D.}~\bibnamefont {Esteve}}, \ and\ \bibinfo {author} {\bibfnamefont {M.~H.}\
  \bibnamefont {Devoret}},\ }\href {\doibase 10.1126/science.1069372}
  {\bibfield  {journal} {\bibinfo  {journal} {Science}\ }\textbf {\bibinfo
  {volume} {296}},\ \bibinfo {pages} {886} (\bibinfo {year}
  {2002})}\BibitemShut {NoStop}%
\bibitem [{\citenamefont {Heinsoo}\ \emph {et~al.}(2018)\citenamefont
  {Heinsoo}, \citenamefont {Andersen}, \citenamefont {Remm}, \citenamefont
  {Krinner}, \citenamefont {Walter}, \citenamefont {Salath\'{e}}, \citenamefont
  {Gasparinetti}, \citenamefont {Besse}, \citenamefont {Poto\v{c}nik},
  \citenamefont {Wallraff},\ and\ \citenamefont {Eichler}}]{Heinsoo2018}%
  \BibitemOpen
  \bibfield  {author} {\bibinfo {author} {\bibfnamefont {J.}~\bibnamefont
  {Heinsoo}}, \bibinfo {author} {\bibfnamefont {C.~K.}\ \bibnamefont
  {Andersen}}, \bibinfo {author} {\bibfnamefont {A.}~\bibnamefont {Remm}},
  \bibinfo {author} {\bibfnamefont {S.}~\bibnamefont {Krinner}}, \bibinfo
  {author} {\bibfnamefont {T.}~\bibnamefont {Walter}}, \bibinfo {author}
  {\bibfnamefont {Y.}~\bibnamefont {Salath\'{e}}}, \bibinfo {author}
  {\bibfnamefont {S.}~\bibnamefont {Gasparinetti}}, \bibinfo {author}
  {\bibfnamefont {J.-C.}\ \bibnamefont {Besse}}, \bibinfo {author}
  {\bibfnamefont {A.}~\bibnamefont {Poto\v{c}nik}}, \bibinfo {author}
  {\bibfnamefont {A.}~\bibnamefont {Wallraff}}, \ and\ \bibinfo {author}
  {\bibfnamefont {C.}~\bibnamefont {Eichler}},\ }\href {\doibase
  10.1103/PhysRevApplied.10.034040} {\bibfield  {journal} {\bibinfo  {journal}
  {Phys. Rev. Appl.}\ }\textbf {\bibinfo {volume} {10}},\ \bibinfo {pages}
  {034040} (\bibinfo {year} {2018})}\BibitemShut {NoStop}%
\end{thebibliography}%

\end{document}